\documentclass[traditabstract]{aa}

\usepackage{graphicx}
\usepackage{txfonts}
\usepackage{natbib}
\bibpunct{(}{)}{;}{a}{}{,}
\usepackage{array}
\usepackage{xspace}
\usepackage{lscape}
\usepackage{url}
\usepackage{arydshln}

\newcommand{\FeKa}{Fe K\ensuremath{\alpha}\xspace}
\newcommand{\kms}{\ensuremath{\mathrm{km\ s^{-1}}}\xspace}

\newcommand{\redchi}{\ensuremath{\chi _\nu ^2}\xspace}

\newcommand{\xmm}{{\it XMM-Newton}\xspace}
\newcommand{\swift}{{Swift}\xspace}
\newcommand{\exosat}{{EXOSAT}\xspace}
\newcommand{\heao}{{HEAO-1}\xspace}
\newcommand{\rosat}{{\it ROSAT}\xspace}
\newcommand{\iue}{{IUE}\xspace}
\newcommand{\euve}{{EUVE}\xspace}

\newcommand{\integral}{{INTEGRAL}\xspace}
\newcommand{\hst}{{HST}\xspace}
\newcommand{\fuse}{{FUSE}\xspace}
\newcommand{\Flam}{\ensuremath{\mathrm{erg}\ \mathrm{s}^{-1}\ \mathrm{cm}^{-2}\ \AA^{-1}}}

\begin{document}

\title{Multiwavelength campaign on Mrk 509}
\subtitle{IV. Optical-UV-X-ray variability and the nature of the soft X-ray excess}

\author{M. Mehdipour \inst{1} 
\and G. Branduardi-Raymont \inst{1}
\and J. S. Kaastra \inst{2, 3}
\and P. O. Petrucci \inst{4}
\and G. A. Kriss \inst{5, 6}
\and G. Ponti \inst{7}
\and A. J. Blustin \inst{8}
\and \newline S. Paltani \inst{9}
\and M. Cappi \inst{10}
\and R. G. Detmers \inst{2, 3}
\and K. C. Steenbrugge\inst{11, 12}
}

\institute{Mullard Space Science Laboratory, University College London, Holmbury St. Mary, Dorking, Surrey, RH5 6NT, UK \\{\email{missagh.mehdipour@ucl.ac.uk}} 
\and SRON Netherlands Institute for Space Research, Sorbonnelaan 2, 3584 CA Utrecht, the Netherlands
\and Sterrenkundig Instituut, Universiteit Utrecht, P.O. Box 80000, 3508 TA Utrecht, the Netherlands
\and UJF-Grenoble 1 / CNRS-INSU, Institut de Plan\'etologie et d'Astrophysique de Grenoble (IPAG) UMR 5274, F-38041, France
\and Space Telescope Science Institute, 3700 San Martin Drive, Baltimore, MD 21218, USA
\and Department of Physics and Astronomy, The Johns Hopkins University, Baltimore, MD 21218, USA
\and School of Physics and Astronomy, University of Southampton, Highfield, Southampton, SO17 1BJ, UK
\and Institute of Astronomy, University of Cambridge, Madingley Road, Cambridge, CB3 0HA, UK
\and ISDC, Geneva Observatory, University of Geneva, ch. d'\'Ecogia 16, CH-1290 Versoix, Switzerland
\and INAF-IASF Bologna, Via Gobetti 101, I-40129 Bologna, Italy
\and Instituto de Astronom\'ia, Universidad Cat\'olica del Norte, Avenida Angamos 0610, Casilla 1280, Antofagasta, Chile
\and Department of Physics, University of Oxford, Keble Road, Oxford, OX1 3RH, UK}

\date{Received 11 March 2011 / Accepted 10 June 2011}

\abstract{}{}{}{}{}
\abstract
{We present the analysis of \xmm and \swift optical-UV and X-ray observations of the Seyfert-1/QSO \object{Mrk 509}, part of an unprecedented multi-wavelength campaign, investigating the nuclear environment of this AGN. The \xmm data are from a series of 10 observations of about 60 ks each, spaced from each other by about 4 days, taken in Oct-Nov 2009. During our campaign, Mrk 509 was also observed with \swift for a period of about 100 days, monitoring the behaviour of the source before and after the \xmm observations. With these data we have established the continuum spectrum in the optical-UV and X-ray bands and investigated its variability on the timescale of our campaign with a resolution time of a few days. In order to measure and model the continuum as far as possible into the UV, we also made use of Hubble Space Telescope (\hst) Cosmic Origin Spectrograph (COS) observations of Mrk 509 (part of our coordinated campaign) and of an archival Far Ultraviolet Spectroscopic Explorer (\fuse) observation. We have found that in addition to an X-ray power-law, the spectrum displays soft X-ray excess emission below 2 keV, which interestingly varies in association with the thermal optical-UV emission from the accretion disc. The change in the X-ray power-law component flux (albeit smaller than that of the soft excess), on the other hand, is uncorrelated to the flux variability of the soft X-ray excess and the disc component on the probed timescale. The results of our simultaneous broad-band spectral and timing analysis suggest that, on a resolution time of a few days, the soft X-ray excess of Mrk 509 is produced by the Comptonisation of the thermal optical-UV photons from the accretion disc by a warm (0.2 keV) optically thick (${\tau \sim 17}$) corona surrounding the inner regions of the disc. This makes Mrk 509, with a black hole mass of about 1--3 $\times 10^8 \ \mathrm{M}_{\odot}$, the highest mass known system to display such behaviour and origin for the soft X-ray excess.}

\keywords{galaxies: active -- galaxies: nuclei -- galaxies: Seyfert -- quasars: individual: Mrk 509 -- X-rays: galaxies}
\authorrunning{M. Mehdipour et al.}
\titlerunning{Multiwavelength campaign on Mrk 509. IV}
\maketitle

\section{Introduction}
Variability studies of Active Galactic Nuclei (AGN) over the last few decades have been essential in developing our understanding of the physical structure of emitting regions in the vicinity of the central engine. The huge amount of energy radiated away is explained by the current paradigm of accretion of material onto a supermassive black hole (SMBH). The emitting regions in the vicinity of the SMBH are too small to be resolved with current telescopes and therefore clues to the processes going on in the central regions come from analysis of their flux and spectral variability. Multi-wavelength monitoring of the AGN variability is a most effective way to investigate complex phenomena such as the accretion onto a SMBH and provides diagnostics of the processes in the nuclear environment of AGN.

The most prominent feature of AGN Spectral Energy Distributions (SEDs) is the so-called `big-blue-bump' peaking in the extreme UV, which was first attributed to thermal emission from the accretion disc by \citet{Shi78}; it is thought that higher energy photons (UV) originate from small radii in the disc where the temperature is higher, while optical photons are produced further out. The first X-ray spectrum of a Seyfert-1 galaxy, which established the power-law shape of the X-ray continuum, was obtained for NGC 4151, using the proportional counter onboard the {\it Ariel-V} satellite, by \citet{Ive76}. The power-law component in the X-ray spectra of AGN is generally believed to be the result of Compton up-scattering of optical-UV photons from the disc by energetic electrons in a hot corona surrounding it. Apart from the power-law, the X-ray continuum often comprises another component, the soft X-ray excess, which was first discovered in the spectrum of Mrk 509 by \citet{Sin85} using data from the first mission of High Energy Astronomy Observatories (\heao). Shortly afterwards, the presence of a soft excess was also reported in another Seyfert-1 galaxy, \object{Mrk 841}, by \citet{Arn85} using European X-ray Observatory Satellite (\exosat) observations. Since then, the soft excess has been observed in the X-ray spectra of the majority of AGN and its origin remains an area of active research to this date. Different interpretations for the nature of the soft X-ray excess can be found in the literature such as: (1) the high energy tail of the thermal emission from the accretion disc (e.g. \citealt{Arn85}, \citealt{Pou86}); (2) an artefact of strong, relativistically smeared, partially ionized absorption in a wind from the inner disc (e.g. \citealt{Gie04}); (3) part of the relativistically blurred photoionised disc reflection spectrum modelled by \citet{Ros05} (e.g. \citealt{Cru06}); (4) `warm' Comptonisation: up-scattering of seed disc photons in a Comptonising medium which has lower temperature and higher optical depth than the corona responsible for the X-ray power-law emission (e.g. the study by \citealt{Mag98} of the Seyfert-1 \object{NGC 5548}; the analyses by \citealt{Mid09} and \citealt{Jin09} of two Narrow Line Seyfert-1s (NLS1s): RE J1034+396 and RX J0136.9-3510).

Past multi-wavelength monitoring campaigns of Seyfert-1 galaxies have shown striking similarities in the optical-UV and X-ray variability, suggesting a strong link between the emission in these energy ranges, such as that provided by inverse Comptonisation. \citet{Wal93} studied the soft X-ray (0.1--2.4 keV) spectra of 58 Seyfert-1 AGN, including Mrk 509, using \rosat observations. They reported presence of a soft X-ray excess above the extrapolation of the hard X-ray power-law in 90\% of the sources. They also found that the soft X-ray excess is well correlated to the strength of the big-blue-bump observed with the International Ultraviolet Explorer (\iue) in the UV band. They concluded that the big-blue-bump is an ultraviolet to soft X-ray feature, which has a similar shape in all Seyfert-1 galaxies of their sample. From an intensive multi-instrument campaign of \object{NGC 4151}, \citet{Ede96} found that optical-UV and 1--2 keV X-ray fluxes varied together on timescale of days, with the X-ray flux varying with much larger amplitude than the optical-UV; the phase difference between UV and 1--2 keV X-ray was consistent with zero lag, with an upper limit of $\lesssim 0.3$ days. \citet{Mar97} measured the spectrum and lightcurve of \object{NGC 5548} over a period of 2 months using the Extreme Ultraviolet Explorer (\euve) when the galaxy was also monitored with \hst. They found the optical-UV and EUV variations to be simultaneous with the amplitude in the EUV twice that in the UV. Also the shape of the \euve spectrum was consistent with a gradual decreasing of flux from the UV through to the soft X-ray with no emission lines detected in the EUV band. Furthermore, from a monitoring study of Mrk 509 using Rossi X-ray Timing Explorer (RXTE) and ground-based optical observations, \citet{Mar08} report a strong correlation between variability of the optical and hard X-ray emission on timescale of a few years, with optical variations leading the hard X-rays by about 15 days. 

The work presented here is part of a large multi-wavelength campaign studying the nearby Seyfert-1/QSO Mrk 509. This AGN has a cosmological redshift of 0.034397 (\citealt{Huc93}) corresponding to a luminosity distance of 145 Mpc (taking ${H_{0}=73\ \mathrm{km\ s^{-1}\ Mpc^{-1}}}$, $\Omega_{\Lambda}=0.73$ and $\Omega_{m}=0.27$). The details of our multi-wavelength campaign are presented in \citet{Kaa11}. As part of this campaign, \swift monitoring was carried out before and after a series of observations by \xmm and the International Gamma-Ray Astrophysics Laboratory (\integral). One of the aims of this campaign is determining the location and structure of the multi-phase warm absorber outflows in Mrk 509, which is done by monitoring the response of the warm absorber to intrinsic source variability using high resolution spectra from the Reflection Grating Spectrometer (RGS) onboard  \xmm. In this work we focus particularly on the optical-UV and X-ray source variability of Mrk 509 using data from the \xmm Optical Monitor (OM) and the European Photon Imaging Camera (EPIC-pn). Our goal in this paper is to try and explain the intrinsic multi-wavelength variability of Mrk 509 and its relation to the soft X-ray excess in terms of physical changes in the accretion disc and corona. 

The structure of this paper is as follows. In Section 2 we describe the observations and data reduction, Sect. 3 focuses on the photometric and spectroscopic corrections applied to the data in order to establish the continuum; the spectral modelling is described in detail in Sects. 4 and 5; we discuss our findings in Sect. 6 and give concluding remarks in Sect. 7.

\section{Observations and data analysis}
Some details of the \xmm's OM and EPIC-pn observations and of the \swift's UV/Optical Telescope (UVOT) and X-Ray Telescope (XRT) observations are shown in Table \ref{obs_table}.

\begin{table*}
\begin{minipage}[t]{\hsize}
\setlength{\extrarowheight}{3pt}
\caption{The \xmm (OM, EPIC-pn) and \swift (UVOT, XRT) observations details.}
\label{obs_table}
\centering
\renewcommand{\footnoterule}{}
\begin{tabular}{l l l l l | c c c c c c c | c c}
\hline \hline
 & & & \multicolumn{2}{c|}{Start time (UTC)} & \multicolumn{7}{c|}{Total exposure time at each filter (ks)} & \multicolumn{2}{c}{X-ray exposure (ks) \footnote{Cleaned X-ray exposure times are shown.}} \\
Inst. & Obs. & ID & yyyy-mm-dd & hh:mm:ss & V & B & U & W1 & M2 & W2 & Vgrism & EPIC-pn & XRT \\ 
\hline

OM & 1 & 0601390201 & 2009-10-15 & 07:16:25 & 6.0 & 7.5 & 7.5 & 7.5 & 7.5 & 8.1 & 5.0 & 40.0 & - \\
OM & 2 & 0601390301 & 2009-10-19 & 15:47:02 & 7.5 & 6.0 & 3.0 & 7.5 & 8.0 & 8.7 & 5.0 & 44.4 & - \\
OM & 3 & 0601390401 & 2009-10-23 & 06:08:12 & 6.0 & 7.5 & 7.5 & 7.5 & 7.5 & 8.1 & 5.0 & 42.4 & - \\
OM & 4 & 0601390501 & 2009-10-29 & 07:22:07 & 7.5 & 7.5 & 7.5 & 7.5 & 6.0 & 4.9 & 5.0 & 42.4 & - \\
OM & 5 & 0601390601 & 2009-11-02 & 03:12:58 & 7.5 & 7.5 & 7.5 & 7.5 & 7.9 & 7.8 & 5.0 & 43.7 & - \\
OM & 6 & 0601390701 & 2009-11-06 & 07:27:14 & 7.5 & 7.5 & 7.5 & 7.5 & 7.5 & 8.5 & 5.0 & 43.9 & - \\
OM & 7 & 0601390801 & 2009-11-10 & 09:08:51 & 7.5 & 7.5 & 7.5 & 7.5 & 7.5 & 5.0 & 5.0 & 42.3 & - \\
OM & 8 & 0601390901 & 2009-11-14 & 08:53:43 & 7.5 & 7.5 & 6.0 & 7.5 & 7.5 & 6.3 & 5.0 & 42.3 & - \\
OM & 9 & 0601391001 & 2009-11-18 & 02:35:08 & 7.5 & 7.5 & 7.5 & 7.5 & 8.9 & 9.5 & 5.0 & 45.6 & - \\
OM & 10 & 0601391101 & 2009-11-20 & 08:07:05 & 7.5 & 7.5 & 7.5 & 7.5 & 7.5 & 8.2 & 5.0 & 43.7 & - \\
\hline
UVOT & 1 & 00035469005 & 2009-09-04 & 14:10:14 & - & 0.1 & 0.1 & 0.3 & - & 0.3 & -  & - & 0.9 \\
UVOT & 2 & 00035469006 & 2009-09-08 & 04:06:41 & 0.1 & 0.1 & 0.1 & 0.2 & 0.3 & 0.5 & - & - & 1.3 \\
UVOT & 3 & 00035469007 & 2009-09-12 & 18:58:17 & - & - & - & - & 0.4 & - & - & - & 0.4 \\
UVOT & 4 & 00035469008 & 2009-09-16 & 20:48:13 & - & - & - & - & 1.0 & - & - & - & 1.0 \\
UVOT & 5 & 00035469009 & 2009-09-20 & 07:53:27 & - & - & - & - & - & - & - & - &1.1 \\
UVOT & 6 & 00035469010 & 2009-09-24 & 10:47:43 & - & - & - & - & 1.3 & - & - & - & 1.2 \\
UVOT & 7 & 00035469011 & 2009-10-02 & 18:55:48 & - & - & - & - & 1.0 & - & - & - & 1.0 \\
UVOT & 8 & 00035469012 & 2009-10-06 & 00:53:57 & - & - & - & - & 1.2 & - & - & - & 1.0 \\
UVOT & 9 & 00035469013 & 2009-10-10 & 22:14:30 & - & - & - & - & 1.1 & - & - & - & 1.1 \\
UVOT & 10 & 00035469014 & 2009-10-14 & 11:05:50 & - & - & - & - & 1.3 & - & - & - &  1.4 \\
UVOT & 11 & 00035469015 & 2009-10-18 & 06:05:58 & - & - & - & - & 1.0 & - & - & - & 0.9 \\
UVOT & 12 & 00035469016 & 2009-11-20 & 07:30:53 & - & - & - & 1.0 & - & - & - & - & 1.0 \\
UVOT & 13 & 00035469017 & 2009-11-24 & 23:51:24 & - & - & 1.0 & - & - & - & - & - & 1.0 \\
UVOT & 14 & 00035469018 & 2009-11-28 & 20:10:30 & - & - & - & 1.2 & - & - & - & - & 1.2 \\
UVOT & 15 & 00035469019 & 2009-12-02 & 01:16:34 & - & - & - & 1.3 & - & - & - & - & 1.3 \\
UVOT & 16 & 00035469020 & 2009-12-06 & 07:19:20 & - & - & - & 1.1 & - & - & - & - & 1.1 \\
UVOT & 17 & 00035469021 & 2009-12-08 & 02:42:06 & - & - & - & - & - & 1.1 & - & - & 1.1 \\
UVOT & 18 & 00035469022 & 2009-12-10 & 12:44:41 & 0.6 & - & - & - & 0.5 & - & - & - & 1.1 \\
UVOT & 19 & 00035469023 & 2009-12-12 & 20:41:00 & - & - & - & - & 0.9 & - & - & - & 1.0 \\
\hline
\end{tabular}
\end{minipage}
\end{table*}
%

\subsection{OM and UVOT broad-band filters data}
\label{om_uvot_reduction}

The OM and UVOT data from Image-mode operations were taken with six broad-band filters: V, B, U, UVW1, UVM2 and UVW2. The OM Image mode data were processed with the SAS v9.0 $\mathtt{omichain}$ pipeline. We performed aperture photometry on each image in a fully interactive way using the $\mathtt{omsource}$ program. We selected the source and background regions to extract the count rates, and applied all the necessary corrections, i.e. for the point spread function (PSF) and coincidence losses, including time-dependent sensitivity (TDS) corrections. The OM count rates were extracted from a circle of 12 pixels radius (5.8\arcsec) centred on the source nucleus. The background was extracted from a source-free region of the same radius. Similarly, for the UVOT, we used the $\mathtt{uvotsource}$ tool to perform photometry using the recommended circular aperture radius of 10 pixels (5.0\arcsec) and applying the standard instrumental corrections and calibration according to \citet{Poo08}. For the purpose of spectral fitting with the {\sc spex} code \citep{Kaa96} version 2.01.05\footnote{http://www.sron.nl/spex}, we used the filter count rates and the corresponding response matrices, which were constructed using the effective area of the instruments for each of the filters. To calculate the flux density at the filter effective wavelength from the count rate, we used the conversion factors given in \citet{Tal08} and \citet{Poo08} for the OM and the UVOT respectively. Note that these conversion factors have been obtained from observations of standard stars with known spectral shape; therefore the flux of Mrk 509 needs to be corrected for the presence of strong AGN emission lines in the filter bandpasses. We have thus taken into account the contribution to each filter bandpass of all significant emission lines and of the small-blue-bump (see Sect. \ref{BLR_model_sect} for the OM grism modelling) so that the flux at the effective wavelength of the filter corresponds only to the intrinsic continuum emission.

\begin{figure}[!]
\centering
\resizebox{\hsize}{!}{\includegraphics[angle=0]{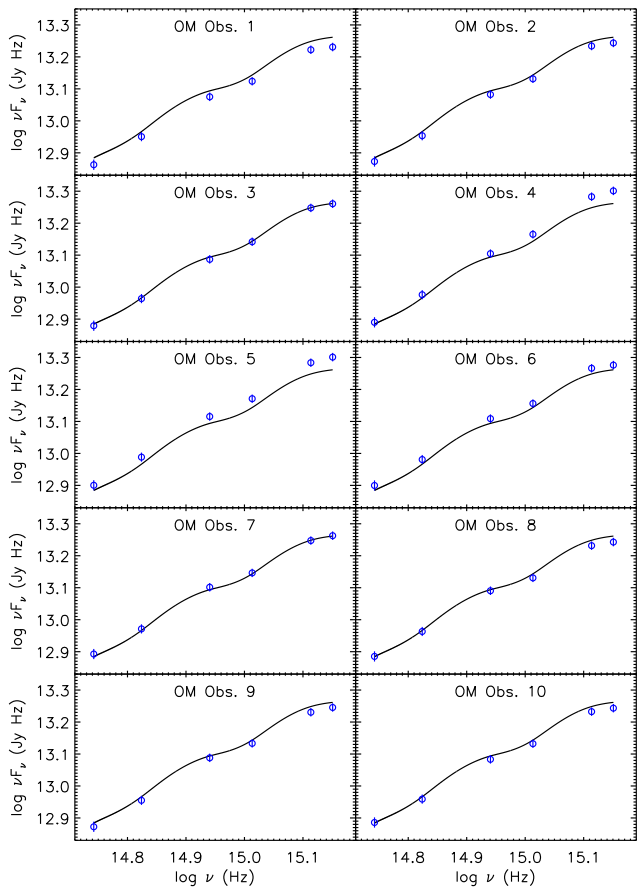}}
\caption{The OM SED data points (shown as open circles) for each observation. The vertical error bars which represent statistical and systematic uncertainties at $1\sigma$ level are mostly within the size of the data symbols. The average spline SED (black curve) obtained from fitting the 10 OM observations is plotted in each panel for reference to show the variability.}
\label{om_SEDs}
\end{figure}

\begin{figure*}[!]
\centering
\resizebox{14.3 cm}{!}{\includegraphics[angle=0]{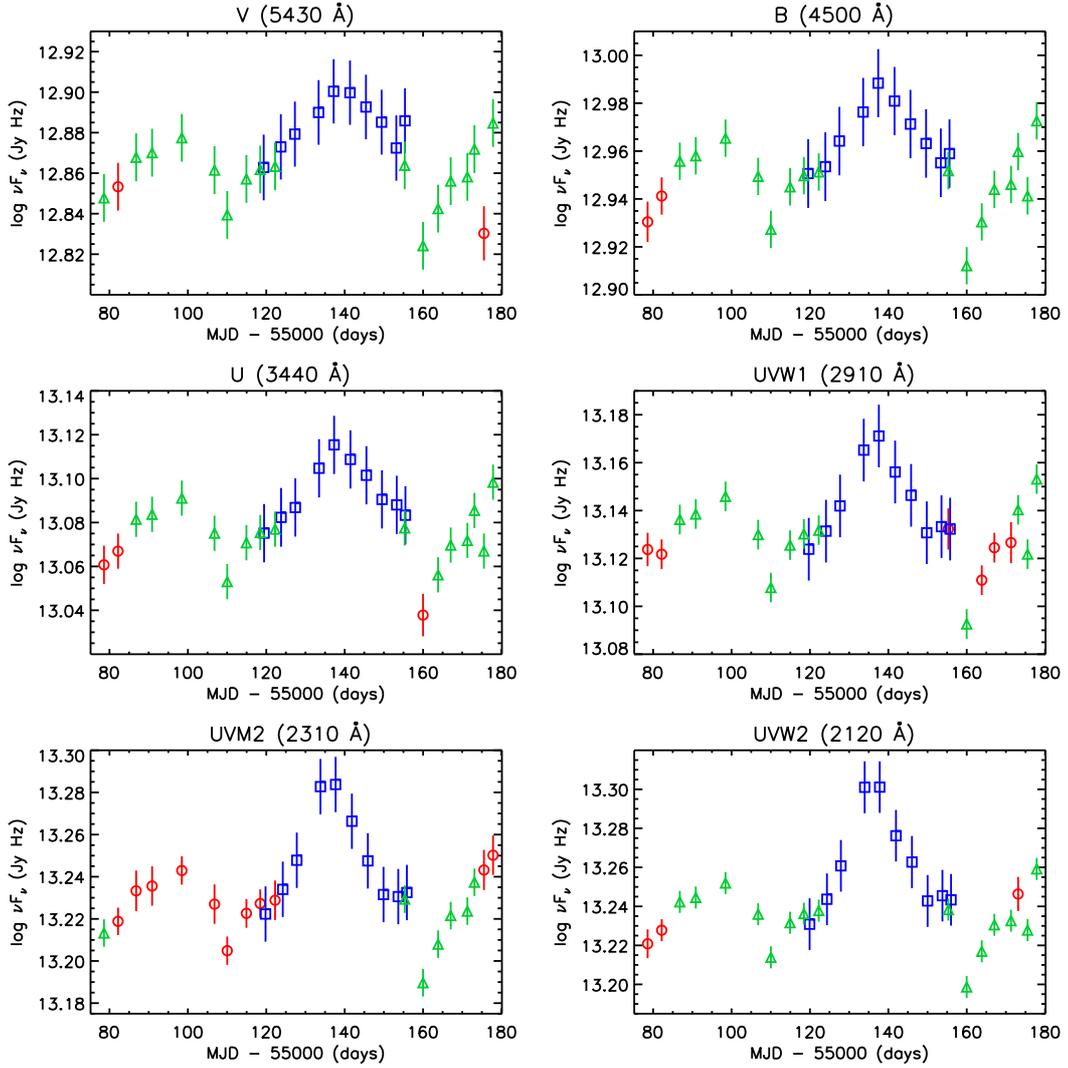}}
\caption{Intrinsic optical-UV continuum lightcurves of Mrk 509 over 100 days at the end of 2009 at the wavelengths indicated above each panel. Blue squares: OM; red circles: measured UVOT; green triangles: UVOT observations for which fluxes have been calculated using a time-averaged SED as described in Sect. \ref{om_uvot_reduction}. The error bars of all lightcurves represent statistical and systematic calibration uncertainties at $1\sigma$ level.}
\label{uvot_om_lightcurves}
\end{figure*}

Next, in order to account for the difference in the calibration of the OM and UVOT instruments, we performed the following to normalise the UVOT fluxes against the OM for each filter. As shown in Table \ref{obs_table}, we have 10 OM observations with exposures in each of the 6 filters. We thus obtained 10 SEDs with 6 optical-UV data points (see Fig. \ref{om_SEDs}) for the source, after implementing all the corrections described in Sect. \ref{corrections_sect}. We then performed a cubic spline interpolation between the data points for each SED and thus obtained a flux versus wavelength relation for each OM observation. During the monitoring campaign there were two occasions when the OM and UVOT observations overlapped (or were very close to overlap). They correspond to OM Obs. 1 and UVOT Obs. 10 (filter M2), and OM Obs. 10 and UVOT Obs. 12 (filter W1). Using the OM SEDs of Obs. 1 and Obs. 10, we calculated the expected OM flux at the effective wavelengths of the UVOT M2 and W1 filters. Then by taking the ratio of the OM and UVOT fluxes, normalisation factors for the M2 and W1 filters of OM and UVOT were obtained. In order to calculate the normalisation factor for the other filters we made use of UVOT Obs. 2 in which all 6 filters were used. By knowing the average shape of the SED over the duration of the 10 OM observations (see Fig. \ref{om_SEDs}) we can scale the SED according to the normalised M2 and W1 fluxes of UVOT Obs. 2. These two SEDs can then be used to give us the normalisation factors of the remaining filters assuming the shape of the average OM SED remains the same over the observations. We find that at the same wavelengths, the OM fluxes are generally larger than the UVOT ones by about 10\% due to calibration differences. The time averaged SED also provides us with a template that can be used to calculate the flux at any particular wavelength for those UVOT observations missing exposures with some of the filters, by knowing only the flux in one of the filters. This assumes that the shape of the average OM SED is a good representation of the source and did not change during the \swift observations, which are before and after the 37-day time span of the OM observations. So finally, the lightcurves in Fig. \ref{uvot_om_lightcurves} show how the OM and UVOT continuum fluxes measured at the same wavelengths (effective wavelength of the OM filters, shown above each panel in Fig. \ref{uvot_om_lightcurves}) varied over the combined 100 days duration of the \xmm and \swift observations.

\subsection{OM optical grism data}
\label{grism_data_sect}

The images from the OM optical grism, effectively a grating-prism combination, were first processed with the SAS v9.0 $\mathtt{omgchain}$ pipeline (for an overview of the OM instrument see \citealt{Mas01}). All the necessary corrections, including that for Modulo-8 fixed pattern noise and removal of scattered light features, were applied to obtain undistorted and rotated grism images. We then used the $\mathtt{omgsource}$ program to interactively identify the zero and first dispersion order spectra of our source, to properly define the source and background extraction regions and extract the calibrated spectrum from the grism images of each observation. The flux calibration by default is performed between 3000 and 6000 $\AA$ for the optical grism in the SAS software. However, because we do detect the $\mathrm{H}\alpha\, \lambda 6563$ emission line right at the long wavelength end of the first order spectrum, we extended the flux calibration to around 7000 $\AA$ in order to include this emission line in our spectra. We used multiple OM spectra of the standard stars GD153 and HZ2 to extend the calibration by normalising them to \hst spectra taken from the {\sc calspec}\footnote{http://www.stsci.edu/hst/observatory/cdbs/calspec.html} database. We find the contamination from the second order spectrum (which overlaps the first order one at the location of the $\mathrm{H}\alpha$ in the image) to be negligible compared with the strong emission from the line. 

\subsection{\hst/COS and archival \fuse data}
In order to measure the continuum as far possible into the UV band (up to the Galactic Lyman limit), we used \hst/COS and \fuse spectra of Mrk 509. The COS observations, part of our coordinated Mrk 509 campaign, were taken on 10 and 11 December 2009 over 10 \hst orbits. The archival \fuse spectrum of Mrk 509 is from 15 October 2000. Since both COS and \fuse share a common wavelength band from about 1165 to 1185 $\AA$, the archival \fuse spectrum was easily scaled to the flux level at the time of the COS observations, assuming the continuum shape in the \fuse band did not vary. The data reduction procedures and the full analysis of the COS spectra are reported in \citet{Kri10}. In this paper for our modelling we use measurements of the UV continuum (corrected for Galactic extinction) from narrow spectral bands, which are free of emission and absorption lines (see Table \ref{cos_fuse_table}). The COS data processing pipeline uses a sensitivity function and the \fuse pipeline uses effective area curves to produce flux-calibrated spectra. Therefore for the purpose of fitting the UV spectra and the optical filter data together within the {\sc spex} package, we had to convert the UV fluxes back to counts per second by changing to photon units and then multiplying by the effective area integrated over the narrow bandpasses given in Table \ref{cos_fuse_table}. The COS observations were made close in time to the UVOT Obs. 18, and since UVOT Obs. 18 is close in flux level to that of OM Obs. 2, we included the COS and the \fuse data when modelling the \xmm Obs. 2 data. For all other \xmm observations, we scaled the COS and \fuse data based on the flux level observed in the OM optical-UV filters.

\begin{table}
\caption{The \fuse and \hst/COS continuum values corrected for Galactic extinction. The \fuse observations were taken on 15 October 2000 and the COS observations are from 10-11 December 2009. The \fuse values have been normalised to COS level.}
\label{cos_fuse_table}
\setlength{\extrarowheight}{3pt}
\begin{minipage}[t]{\hsize}
\renewcommand{\footnoterule}{}
\centering
\begin{tabular}{c c c c}
\hline \hline
 & Band & $F_{\lambda}$ & \\
Instrument & $(\AA)$ & $(\mathrm{erg}\ \mathrm{cm}^{-2}\ \mathrm{s}^{-1}\ \AA^{-1})$ & Flux error \\
\hline

\fuse & $941-944$ & $2.15 \times 10^{-13}$ & $10\%$ \\

\fuse & $959-961$ & $1.93 \times 10^{-13}$ & $10\%$ \\

\fuse & $983-985$ & $1.87 \times 10^{-13}$ & $10\%$ \\

\fuse & $992-994$ & $1.72 \times 10^{-13}$ & $10\%$ \\

\fuse/COS & $1165-1185$ & $1.68 \times 10^{-13}$ & $5\%$ \\

COS & $1405-1425$ & $1.47 \times 10^{-13}$ & $5\%$ \\

COS & $1740-1760$ & $1.07 \times 10^{-13}$ & $5\%$ \\

\hline
\end{tabular}
\end{minipage}
\end{table}

\subsection{EPIC-pn and XRT spectra}
The EPIC-pn data were reduced, starting from the ODF files, using the standard SAS v9.0 software. During the \xmm monitoring, the EPIC-pn camera was operating in the Small-Window mode with the thin-filter applied. We only selected events with $\mathtt{FLAG==0}$ and $\mathtt{\#XMMEA\_EP}$ attributes and screened for increased activity of background particles. During the campaign no strong soft proton flares have been detected. The cleaned EPIC-pn exposure of each \xmm observation is about 60 ks. Nevertheless, the EPIC-pn camera, when operating in the Small-Window mode, is only active for about 70\% of the time. Thus the final cleaned exposure time of each EPIC-pn spectrum is about 40 ks (see Table \ref{obs_table}). The spectra were extracted from a circular region of $40''$ radius centred on the source. The background was extracted from a nearby source-free region of the same size on the same chip. The EPIC-pn data showed no evidence of significant pile-up, thus single and double events were selected. Response matrices were generated for each source spectrum using the SAS tasks $\mathtt{arfgen}$ and $\mathtt{rmfgen}$. More details on the data reduction and analysis of the EPIC data are presented in Ponti et al. (in prep.). For the XRT data reduction, we used the online \swift-XRT data products generator facility with the default settings (details given in \citealt{Eva09}) to produce the X-ray spectra.

\section{Intrinsic continuum emission}
\label{corrections_sect}
In order to establish the intrinsic continuum emission of Mrk 509, we applied photometric and spectroscopic corrections to our data to take into account the processes which take place in our line of sight towards the central engine. Sects. \ref{reddening_sec}, \ref{host_gal_sect} and \ref{BLR_model_sect} are applicable to the optical-UV (OM and UVOT), and Sects. \ref{Gal_abs_sect} and \ref{warm_abs_sect} to the X-ray (EPIC-pn and XRT). In the latter case the presence of Galactic and AGN warm absorption is taken into account during the fitting process. 

%
\subsection{Galactic interstellar dereddening}
\label{reddening_sec}
To correct the OM and UVOT optical-UV fluxes for interstellar reddening in our Galaxy we used the reddening curve of \citet{Car89}, including the update for near-UV given by \citet{ODo94}. The colour excess $E(B-V) = 0.057\ \mathrm{mag}$ is based on calculations of \citet{Sch98} as shown in the NASA/IPAC Extragalactic Database (NED). The scalar specifying the ratio of total selective extinction $R_V \equiv A_V/E(B-V)$ was fixed at 3.1.

%
\subsection{Host galaxy stellar emission}
\label{host_gal_sect}
To correct for the host galaxy starlight contribution in the OM and UVOT fields we have used the results of \citet{Ben09} and \citet{Kin96}. \citet{Ben09} have determined the host galaxy observed flux at the rest-frame wavelength of $5100\ \AA$ ($F_{\mathrm{gal},\ 5100\ \AA}$) for a sample of AGN using \hst images. For Mrk 509, $F_{\mathrm{gal},\ 5100\ \AA} = (2.52 \pm 0.23)\ 10^{-15}\ \Flam$ for an aperture of $5.0\arcsec \times 7.6\arcsec$. In order to calculate the host galaxy spectrum in the optical band, we used a model spectrum and scaled it based on the $F_{\mathrm{gal},\ 5100\ \AA}$ value. Since the OM and UVOT apertures are taking in only the galaxy's innermost few kpc, the galaxy bulge template of \citet{Kin96} was used. Figure \ref{host_spectrum} shows the host bulge model spectrum in the optical band, which becomes gradually negligible going towards the UV.

\citet{Kot94} have also measured the nuclear stellar flux contribution using different broad-band filters in a 6\arcsec diameter circular aperture for a sample of galaxies.  They have done this by subtracting the AGN contribution using profile fitting. The nuclear stellar flux of Mrk 509 for the standard B and V filters are shown in Fig. \ref{host_spectrum}, which are consistent with the host bulge model spectrum.

%
\subsection{Correction for emission from the Broad Line Region (BLR) and Narrow Line Region (NLR)}
\label{BLR_model_sect}
\begin{figure}
\resizebox{\hsize}{!}{\includegraphics[angle=0]{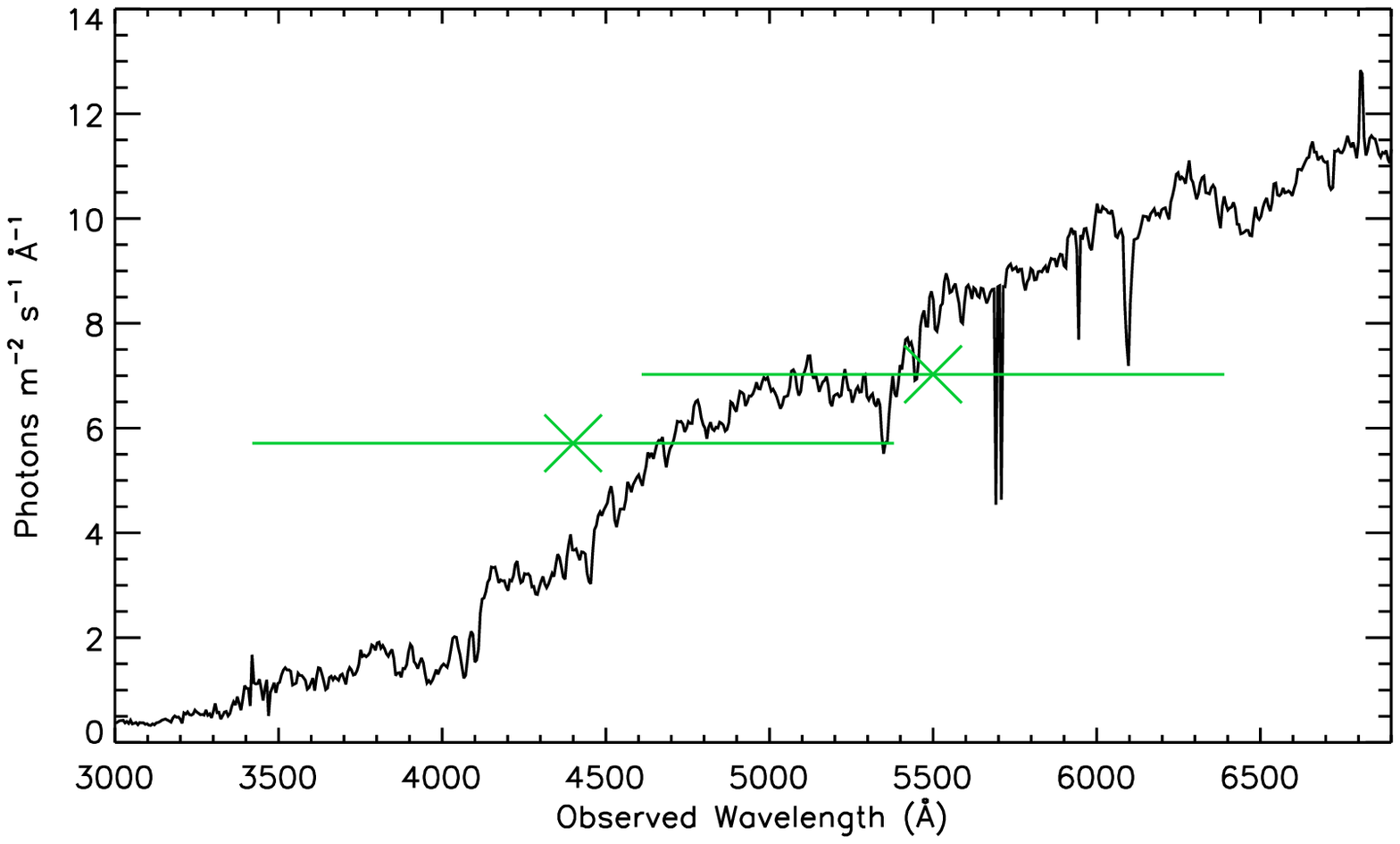}}
\caption{The calculated bulge stellar spectrum of Mrk 509 host galaxy shown in the range observed by the OM optical grism. The spectrum is obtained from the \hst measurement at rest wavelength of $5100\ \AA$ given in \citet{Ben09} and using the bulge model template of \citet{Kin96} to calculate the flux at other wavelengths. The two data points (green crosses) superimposed on the spectrum are the nuclear stellar flux for the standard B and V filters measured by \citet{Kot94}; the horizontal bars represent the filter bandpasses.}
\label{host_spectrum}

\resizebox{\hsize}{!}{\includegraphics[angle=270]{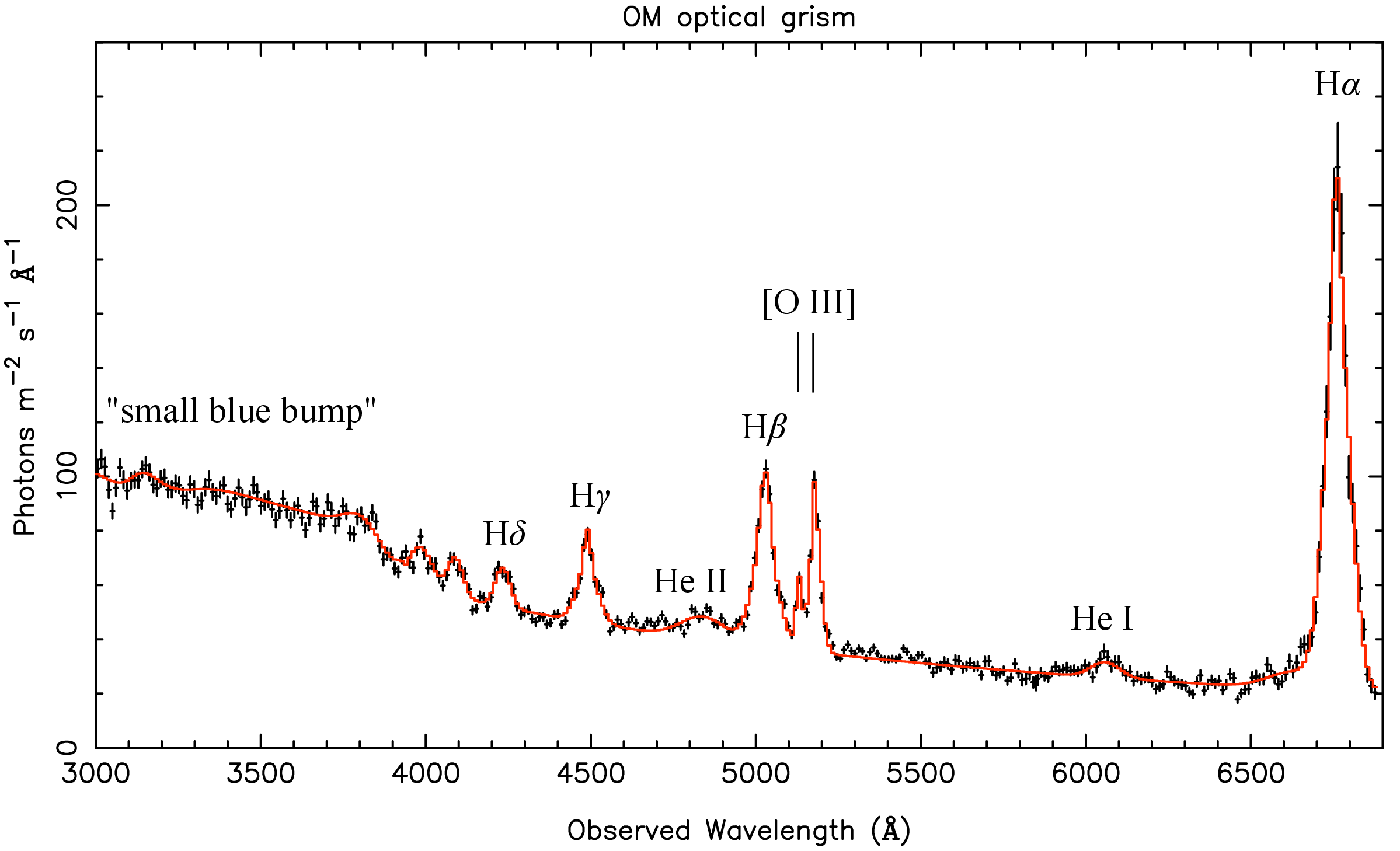}}
\caption{The observed OM optical grism spectrum of Mrk 509 (already corrected for Galactic extinction and the AGN host galaxy stellar emission) obtained by averaging the OM Vgrism spectra from the 10 \xmm observations. The data are shown in black and the model in red. The prominent broad and narrow emission features are labelled. The low-level sinusoidal pattern below 3900 $\AA$ is a residual effect of the Modulo-8 fixed pattern noise.}
\label{grism_spec}
\end{figure}

For each of the 10 \xmm observations we have a 5 ks OM optical grism exposure with which we can model emission from the BLR and NLR. Figure \ref{grism_spec} shows the mean optical grism spectrum of the observations and our model. The accretion disc blackbody model spectrum (used in Sect. \ref{optical_uv_model_sect} and shown for e.g. in Fig. \ref{pn_om_SEDs_corrected}) has a power-law shape in the optical band. Thus, to model the underlying optical continuum (3000--7000 $\AA$) in the OM grism spectra we used a simple power-law rather than disc blackbody model, which requires a wider energy band to be fitted properly. Our main purpose in fitting the OM grism is modelling the emission features and therefore a simple continuum model is preferred (see Sect. \ref{optical_uv_model_sect} for the modelling of the optical-UV continuum with the OM image data). As shown in Fig. \ref{grism_spec} there is excess emission below about 4000 $\AA$ which is likely to be the long-wavelength end of the `small-blue-bump' feature, a blend of Balmer continuum and \ion{Fe}{ii} line emission. Note that the low-level sinusoidal pattern below 3900 $\AA$ in Fig. \ref{grism_spec} is an instrumental effect (residual of the Modulo-8 fixed pattern noise; \citealt{Mas01}), although the overall average flux level is correct. The spectral shape of the small-blue-bump is strongly blended and thus this complex feature was modelled with a cubic spline with linear spacing between 3000 and 4000 $\AA$ (in the observed frame) with 10 grid points. We modelled the broad and narrow emission lines using Gaussian line profiles. The best-fit parameters of the emission features are shown in Table \ref{grism_table}. Then, having obtained a best-fit model for the mean spectrum we applied it to the individual observations and subsequently fitted them. This enables us to look for any changes in the emission lines during the \xmm campaign ($\sim$ 37 days) and see how they compare with the continuum variability. We do not detect any statistically significant variation in the lines like we observe for the continuum (see Fig. \ref{uvot_om_lightcurves}). This is consistent with the results of the monitoring program of Mrk 509 by \citet{Car96} who found the H$\beta$ $\lambda 4861$ and \ion{He}{ii}\ $\lambda 4686$ emission lines to respond to continuum variations with time lags of about 80 and 60 days respectively, considerably longer than the emission line lags measured for other Seyfert galaxies and also longer than the duration of the \xmm campaign. Finally, we proceeded to correct the OM and UVOT filter data (described in Sect. \ref{om_uvot_reduction}) for the emission feature contributions in the different filter bands.

\begin{table}
\caption{Best-fit parameters of the broad and narrow emission features in the mean OM optical grism spectrum. The $\redchi = 1.04$ (302 d.o.f.).}
\label{grism_table}
\setlength{\extrarowheight}{3pt}
\begin{minipage}[t]{\columnwidth}
\renewcommand{\footnoterule}{}
\centering
\begin{tabular}{c c c}
\hline \hline
Emission & Energy flux \footnote{$10^{-13}$ erg $\mathrm{s}^{-1}$ $\mathrm{cm}^{-2}$.} & FWHM \footnote{\kms.} \\
\hline 

H$\alpha$ $\lambda 6563$ & $44 \pm 2$  & $4300 \pm 200$  \\
H$\beta$ $\lambda 4861$ & $17.0 \pm 0.5$ &  $4400 \pm 200$ \\
H$\gamma$ $\lambda 4340$ & $9.0 \pm 0.3$ & $4400 \pm 200$  \\
H$\delta$ $\lambda 4101$ & $4.2 \pm 0.3$ & $3700 \pm 300$ \\
$[$\ion{O}{iii}$]$ $\lambda 4959$  & $2.2 \pm 0.3$ & $1100 \pm 200$ \\
$[$\ion{O}{iii}$]$ $\lambda 5007$ & $7.8 \pm 0.3$ & $1500 \pm 200$ \\
`small-blue-bump' & $44 \pm 5$ \footnote{Calculated between 2900 and 3700 $\AA$ rest-wavelength.} & - \\

\hline
\end{tabular}
\end{minipage}
\end{table}

%
\subsection{Galactic interstellar X-ray absorption correction}
\label{Gal_abs_sect}
In our modelling of the X-ray spectra the effects of the Galactic neutral absorption were included by applying the {\tt HOT} model (collisional ionisation equilibrium) in {\sc spex}. Assuming \citet{Lod03} abundances, the Galactic \ion{H}{i} column density in our line of sight was fixed to $N_{\mathrm{H}}={4.44\times 10^{20}\ \mathrm{cm}^{-2}}$ \citep{Mur96} and the gas temperature to 0.5 eV to mimic a neutral gas.

%
\subsection{Warm absorber correction}
\label{warm_abs_sect}
In order to perform a multi-wavelength intrinsic variability study one also needs to determine the X-ray continuum before absorption by the AGN warm absorber. To this end, we used the {\sc spex} {\tt XABS} photoionised absorption model with the parameters of \citet{Det10b} derived for the stacked 600 ks RGS spectrum gathered during the 2009 campaign. Note that the correction for absorption by the warm absorber is too small to account for the observed X-ray variability. The warm absorption correction is less than 6\% of the X-ray photon flux in all the observations, whereas from Obs. 1 to Obs. 5 the photon flux increases by about 45\%.

\section{Independent modelling of the optical-UV and X-ray continua}
\label{preliminary_sect}
In this section we describe the models used to fit the optical-UV (OM) and X-ray spectra (EPIC-pn) of Mrk 509 separately. In Sect. \ref{compt_section} we perform simultaneous broad-band fits with our final model. Note that all the photometric and spectroscopic corrections described in the previous section, and the cosmological redshift have been taken into account here. All the fitted parameter errors quoted in this paper correspond to a $\Delta \chi ^2$ of 1. Parameters of the models given in the tables correspond to the source reference frame, whereas all the figures in this paper are displayed in the observed frame.

\subsection{The X-ray continuum and \FeKa line}
\label{xray_mod_sect}
\begin{figure}
\centering
\resizebox{\hsize}{!}{\includegraphics[angle=270]{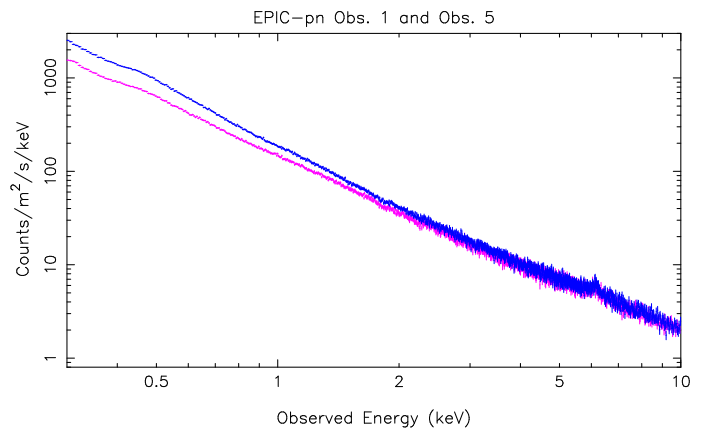}}
\caption{EPIC-pn spectra of Obs. 1 (magenta) and Obs. 5 (blue) binned to a minimum of 100 counts per bin. The spectrum of Obs. 5 has a higher flux than that of Obs. 1 at low energies.}
\label{pn1_pn5}
\end{figure}

\begin{figure}
\centering
\resizebox{\hsize}{!}{\includegraphics[angle=270]{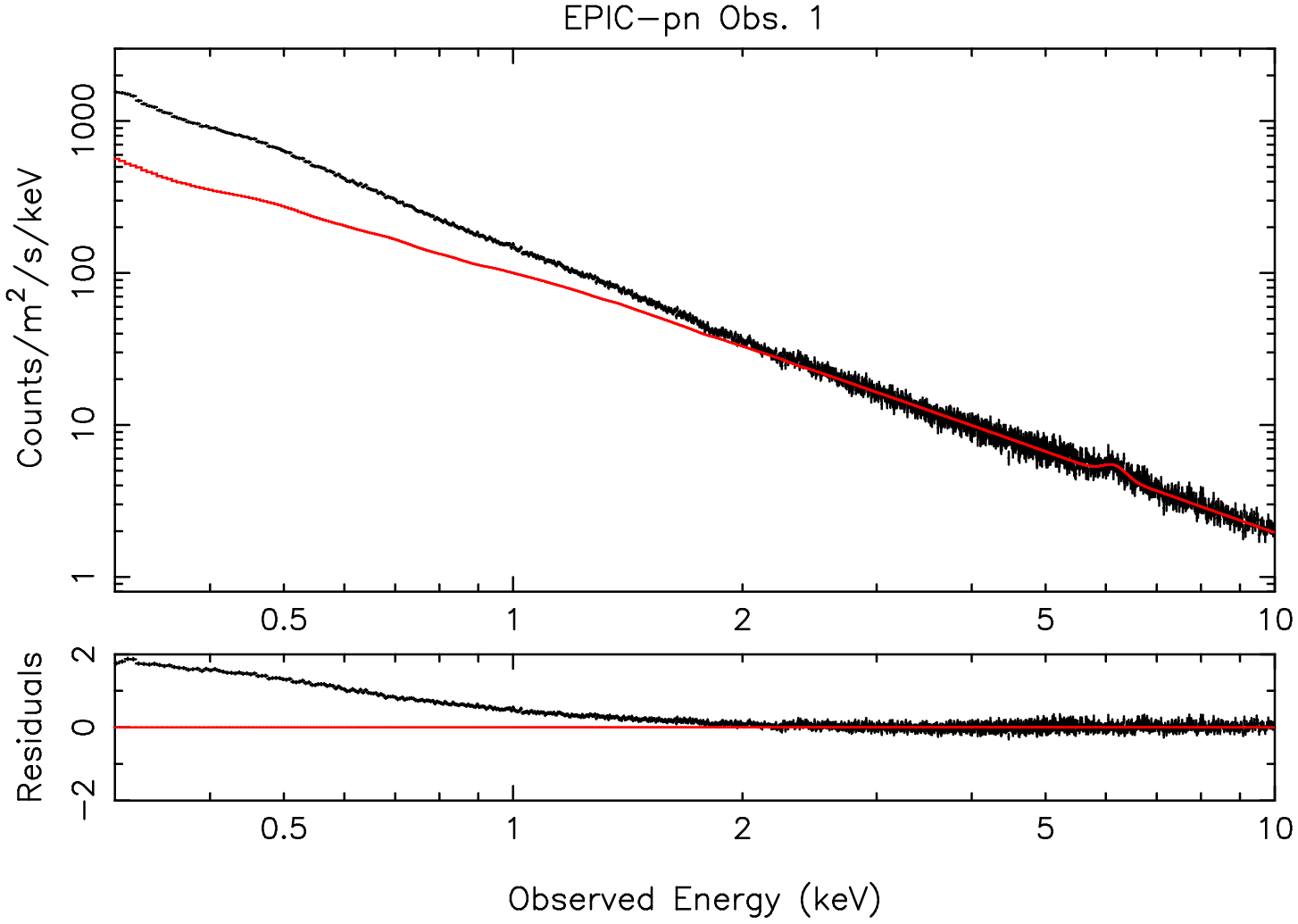}}
\caption{Obs. 1 EPIC-pn single power-law fit (including a Gaussian \FeKa line, Galactic and warm absorption) over the 2.5--10 keV range. The fit is extrapolated to lower energies, displaying the presence of a soft excess below about 2 keV. The data are shown in black and model in red. Residuals of the fit, (Observed$-$Model)$/$Model, are displayed in the bottom panel.}
\label{pn1_v1}
\resizebox{\hsize}{!}{\includegraphics[angle=270]{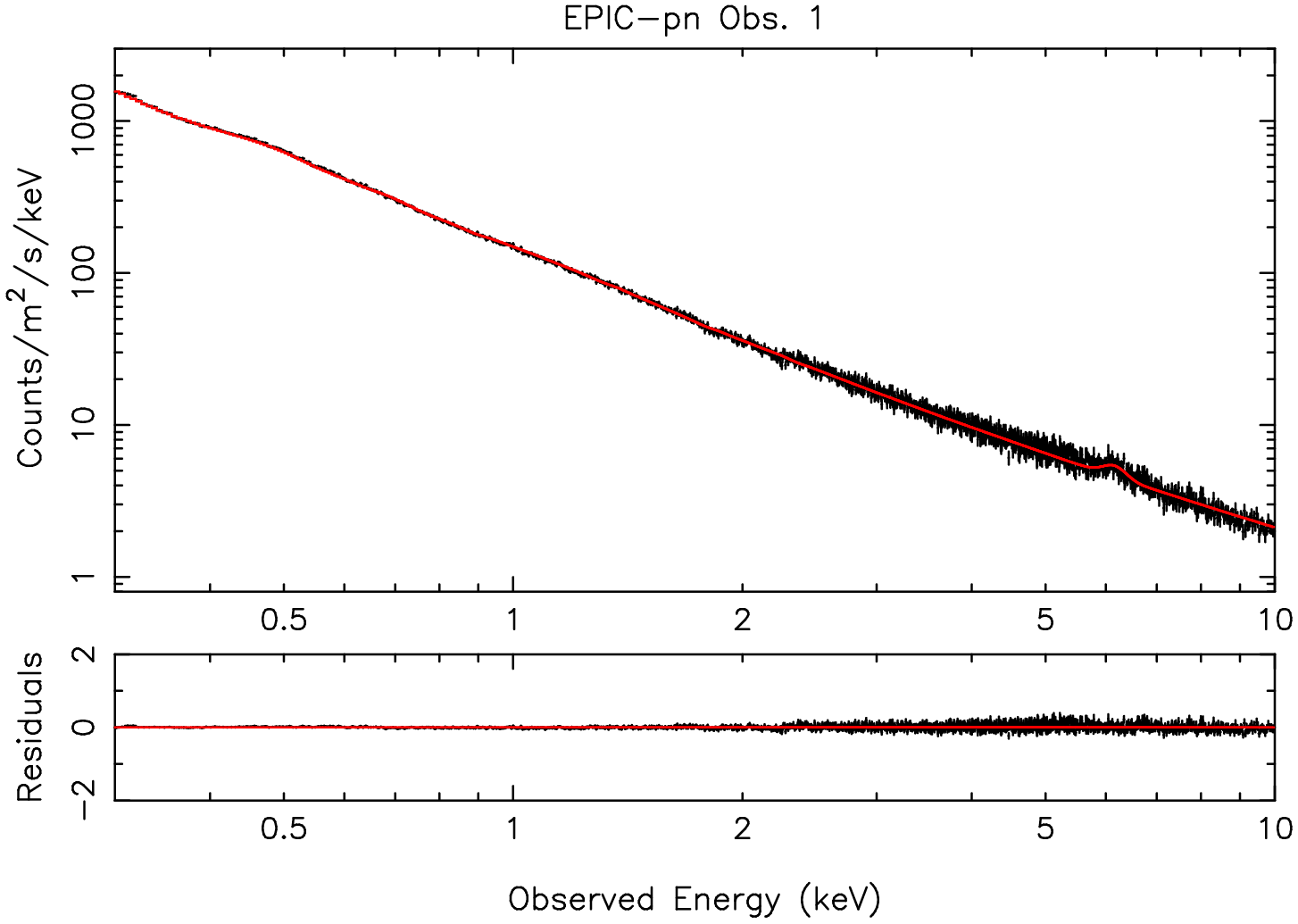}}
\caption{Obs. 1 EPIC-pn two power-law fit (including a Gaussian \FeKa line, Galactic and warm absorption) over the 0.3--10 keV range. The data are shown in black and model in red. Residuals of the fit, (Observed$-$Model)$/$Model, are displayed in the bottom panel.}
\label{pn1_v2}
\end{figure}

The EPIC-pn spectra of Obs. 1 and Obs. 5 are shown together in Fig. \ref{pn1_pn5} to display the extremes of the X-ray variability observed in Mrk 509 during the \xmm campaign. We began our spectral modelling by fitting the EPIC-pn 0.3--10 keV X-ray continuum and the \FeKa line at $\sim$ 6.4 keV of Obs. 1 with a simple power-law ({\tt POW} in {\sc spex}) and a Gaussian line profile ({\tt GAUS} in {\sc spex}) including the absorption models described in Sects. \ref{Gal_abs_sect} and \ref{warm_abs_sect}. We found that a single power-law cannot fit the continuum well between 0.3--10 keV because of the presence of a soft excess below about 2 keV. For a single power-law fit over 0.3--10 keV a $\Gamma$ of 2.27 and a \redchi of 7.9 (1788 d.o.f.) are obtained, whereas for a fit between 2.5--10 keV we find $\Gamma = 1.79$ and $\redchi = 1.01$ (1353 d.o.f.). Figure \ref{pn1_v1} shows the 2.5--10.0 keV EPIC-pn fit for Obs. 1, extrapolated to lower energies, clearly displaying the presence of a steeper continuum below about 2 keV. 
%

%
%

We then fitted the 0.3--10 keV continuum of Obs. 1 by adding a second power-law component to model the soft excess, which resulted in a best-fit with \redchi of 1.09 (1786 d.o.f.). The photon indices of the two power-laws are significantly different: $\Gamma = 2.75$ for the soft X-ray component and $\Gamma = 1.46$ for the hard one. Figure \ref{pn1_v2} shows the two power-law best-fit to the data of Obs. 1. We then applied this model to the spectra of all 10 EPIC-pn observations. Table \ref{cont_table} shows the best-fit parameters for the two power-law model for each observation. 

\begin{table}
\caption{Best-fit parameters of the X-ray continuum, obtained with a two power-law model for the EPIC-pn spectra of each observation as described in Sect. \ref{xray_mod_sect}.}
\label{cont_table}
\setlength{\extrarowheight}{3pt}
\begin{minipage}[t]{\hsize}
\renewcommand{\footnoterule}{}
\centering
\fontsize{8.2}{9.2} \selectfont
\begin{tabular}{c | l l | l l | l}
\hline \hline
 & \multicolumn{2}{c|}{Soft power-law:} & \multicolumn{2}{c|}{Hard power-law:} &  \\ 
Obs. & $\Gamma$ \footnote{Photon index.} & Norm \footnote{$10^{52}$ photons $\mathrm{s}^{-1}$ $\mathrm{keV}^{-1}$ at 1 keV.} & $\Gamma$ $^a$ & Norm $^b$ & $\chi_\nu ^2$ / d.o.f. \\
\hline

1 & $2.75 \pm 0.01$ & $3.4 \pm 0.1$ & $1.46 \pm 0.02$ & $1.4 \pm 0.1$ & $1.09/1786$ \\

2 & $2.68 \pm 0.01$ & $3.8 \pm 0.1$ & $1.41 \pm 0.02$ & $1.3 \pm 0.1$ & $1.07/1822$ \\

3 & $2.75 \pm 0.01$ & $4.7 \pm 0.1$ & $1.42 \pm 0.02$ & $1.4 \pm 0.1$ & $1.22/1817$ \\

4 & $2.97 \pm 0.01$ & $3.6 \pm 0.1$ & $1.47 \pm 0.01$ & $1.4 \pm 0.1$ & $1.17/1780$ \\

5 & $2.95 \pm 0.01$ & $4.7 \pm 0.1$ & $1.46 \pm 0.01$ & $1.6 \pm 0.1$ & $1.17/1804$ \\

6 & $2.77 \pm 0.01$ & $4.5 \pm 0.1$ & $1.45 \pm 0.02$ & $1.7 \pm 0.1$ & $1.18/1838$ \\

7 & $2.75 \pm 0.01$ & $4.5 \pm 0.1$ & $1.45 \pm 0.02$ & $1.5 \pm 0.1$ & $1.17/1822$ \\

8 & $2.63 \pm 0.01$ & $4.8 \pm 0.1$ & $1.40 \pm 0.02$ & $1.4 \pm 0.1$ & $1.13/1831$ \\

9 & $2.76 \pm 0.01$ & $4.1 \pm 0.1$ & $1.42 \pm 0.02$ & $1.3 \pm 0.1$ & $1.27/1821$ \\

10 & $2.71 \pm 0.01$ & $4.2 \pm 0.1$ & $1.44 \pm 0.02$ & $1.6 \pm 0.1$ & $1.11/1840$ \\

\hline
\end{tabular}
\end{minipage}
\end{table}
%

The subject of this paper is the broad-band modelling of the Mrk 509 continuum, so we have modelled the \FeKa line using a simple Gaussian profile which gives a good fit. We omit reporting parameters of this line because they are not relevant to the main thrust of this work, and instead refer the reader to another paper reporting our campaign (Ponti et al. in prep.), which is completely dedicated to the study of the \FeKa line in Mrk 509; the data from our long campaign make it possible to study the profile and variability of the \FeKa band in great detail and deserve a separate publication. From previous \xmm and {\it Suzaku} observations of Mrk 509 reported by \citet{Pon09}, the neutral \FeKa line does not show signs of relativistic reflection and the ionised part of the line is complex. Furthermore, in another forthcoming paper (Petrucci et al. in prep.) the analysis will include the \integral data and a reflection model will be applied to fit the Mrk 509 spectra, as higher energy data can constrain the reflection parameters more tightly.

\begin{figure}
\centering
\resizebox{8.5 cm}{!}{\includegraphics{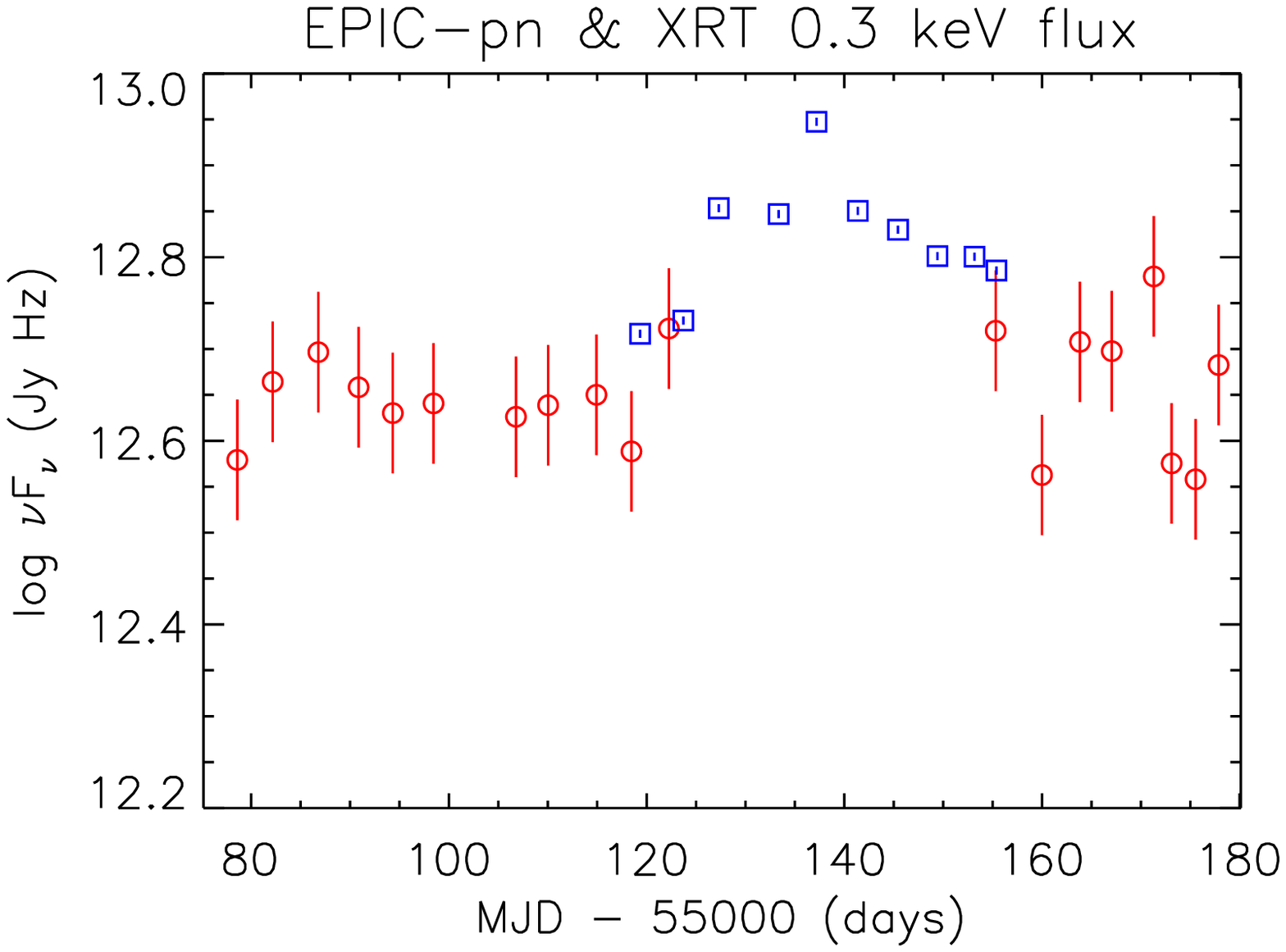}}
\resizebox{8.5 cm}{!}{\includegraphics{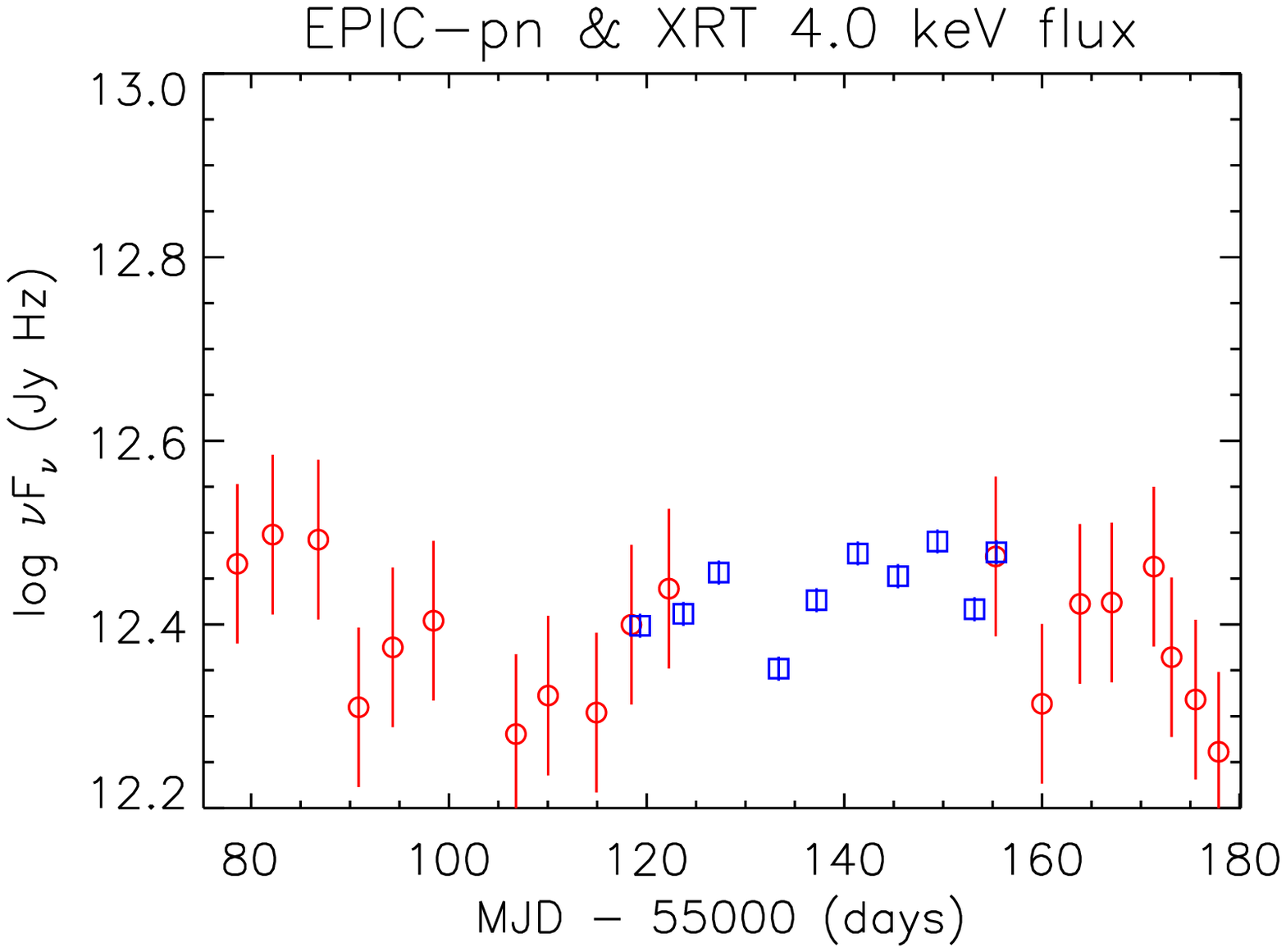}}
\caption{X-ray continuum flux lightcurves of Mrk 509 over 100 days at 0.3 keV (top panel) and 4.0 keV (bottom panel). The red circles (with large error bars) represent the XRT points and the blue squares represent the EPIC-pn points. The bandwidths used to calculate the 0.3 keV fluxes are 0.02 keV for EPIC-pn and 0.3 keV for XRT, and the bandwidths used to calculate the 4.0 keV fluxes are 0.1 keV for EPIC-pn and 2.0 keV for XRT.}
\label{pn_xrt_lightcurve}
\end{figure}

Next, we examined the \swift XRT spectra of Mrk 509 to get a measure of the X-ray flux before and after the \xmm campaign. However, one needs to note that the XRT exposures are very short at around 1 ks (see Table \ref{obs_table}), and also the effective area of the XRT is very much lower than that of EPIC-pn; thus the XRT spectra are of very low quality, and make it impossible to perform any accurate spectral modelling. Therefore, we omit reporting parameters of XRT fits and instead in Fig. \ref{pn_xrt_lightcurve} we show how the soft X-ray (0.3 keV) and hard X-ray (4 keV) continuum fluxes varied for both XRT and EPIC-pn over the 100 days duration of our campaign. 

%

\begin{figure*}[!]
\centering
\resizebox{0.90\hsize}{!}{\includegraphics[angle=0]{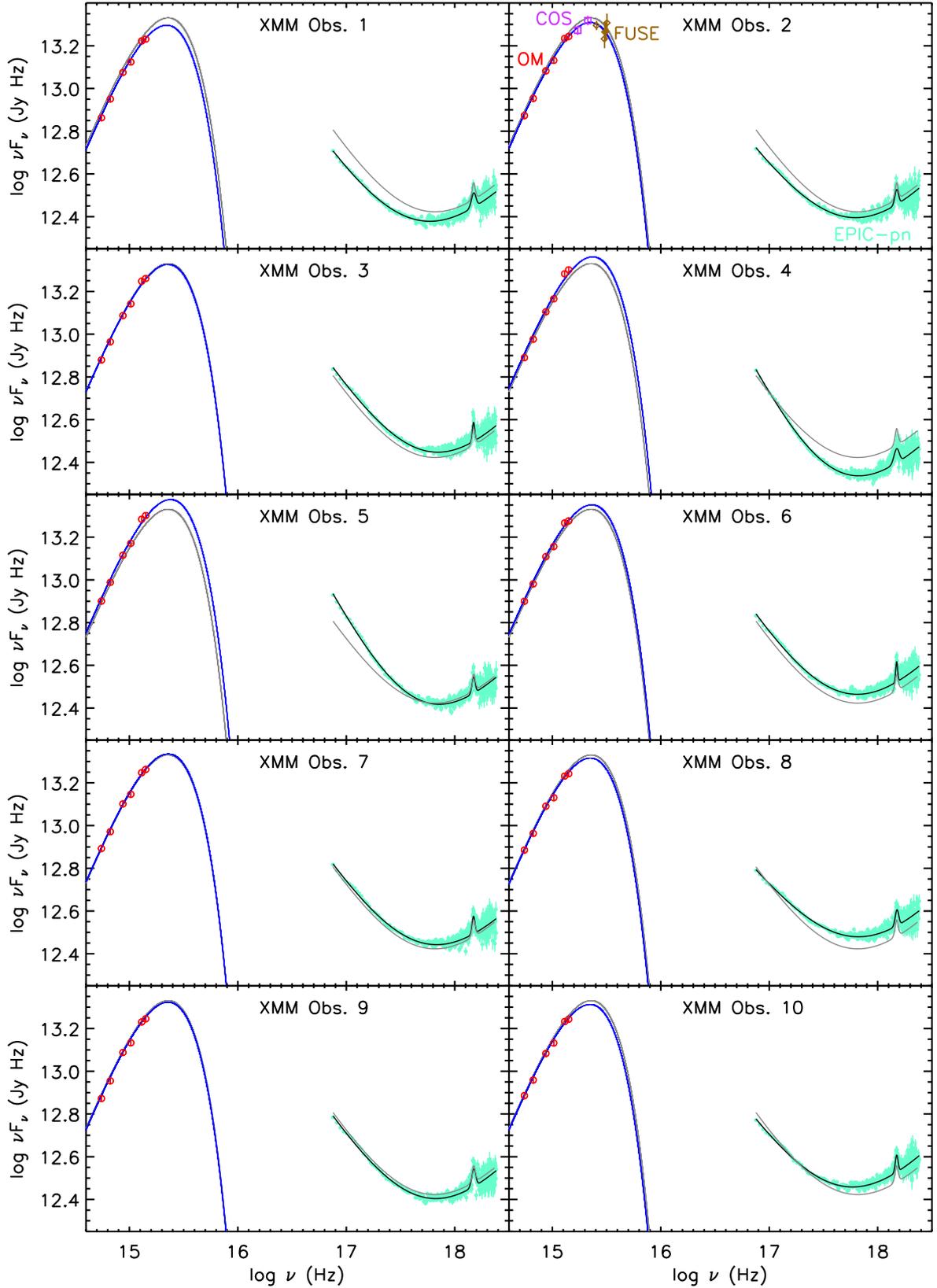}}
\caption{The best-fit model derived in Sect. \ref{preliminary_sect} independently for the optical-UV and X-ray bands for each \xmm observation. The  absorption-corrected EPIC-pn data are show in green and the black lines are the two power-law model fits to the EPIC-pn including a Gaussian for modelling the \FeKa line. The red circles represent the OM data and the blue lines are the disc blackbody models. In panel 2, the purple squares are the \hst/COS data and the brown diamonds are the scaled \fuse data. The average model obtained from the 10 \xmm observations is shown in each panel (in grey) for reference to show the variability.}
\label{pn_om_SEDs_corrected}
\end{figure*}
%
%

\subsection{The optical-UV continuum}
\label{optical_uv_model_sect}
Apart from line emission from the BLR and NLR (which has been taken into account in our modelling of the optical continuum of Mrk 509 - see Sect. \ref{BLR_model_sect}), the dominant feature of the optical-UV spectra of AGN is the big-blue-bump: this was first attributed to thermal emission from the accretion disc by \citet{Shi78}, which has since become the generally agreed paradigm. Annuli of the disc at different radii have different temperatures and therefore the disc spectrum is the weighted sum of different blackbody spectra. To fit the big-blue-bump of Mrk 509, we used the disc blackbody model in {\sc spex} ({\tt DBB}) which is based on a geometrically thin, optically thick Shakura-Sunyaev accretion disc \citep{Sha73}. The radial temperature profile of the accretion disc is given by ${T(r) = \{ 3GM\dot M [ 1 - (R_{{\rm{in}}} /r)^{1/2} ] / (8\pi r^3 \sigma) \}^{1/4}}$ where $T$ is the local temperature at radius $r$ in the disc, $G$ the gravitational constant, $M$ the mass of the SMBH, $\dot M$ the accretion rate, $\sigma$ the Stefan-Boltzmann constant and $R_{{\rm{in}}}$ is the radius at the inner edge of the disc. The flux at frequency $\nu$ from the disc seen by an observer at distance $d$ is given by ${F_\nu   = [4\pi h\nu ^3 \cos i/(c^2 d^2)] \int_{R_{{\rm{in}}} }^{R_{{\rm{out}}} } {({\rm{e}}^{h\nu /kT(r)}  - 1)^{ - 1}\, r\, {\rm{d}}r}}$ where $i$ is the inclination angle of the disc, $R_{{\rm{out}}}$ the radius at the outer edge of the disc, $h$ the Planck constant and $c$ the speed of light (e.g. see \citealt{Fra02}). The parameter $R_{{\rm{out}}}$ cannot be constrained by spectral fitting since most of the optical-UV radiation originates from the inner regions of the disc for a SMBH, thus we fixed $R_{{\rm{out}}}$ to $10^{3} R_{{\rm{in}}}$. The fitted parameters of the model are the normalisation ${A = R_{{\rm{in}}}^2 \cos i}$ and the maximum temperature of the disc $T_{\rm{max}}$, which occurs at $r=(49/36) R_{{\rm{in}}}$.  

Table \ref{uv_cont_table} shows the best-fit parameters of the disc blackbody model for each \xmm observation. The $T_{{\rm{max}}}$ increases from its minimum in Obs. 1 to its maximum in Obs. 5 and then gradually decreases back to the level of Obs. 1 towards the end of the \xmm campaign. The normalisation area of the disc blackbody remains unchanged within errors, which implies that $R_{{\rm{in}}}$ and inclination $i$ parameters have not significantly varied. Thus the variability in the disc temperature comes from a change in the accretion rate. Figure \ref{pn_om_SEDs_corrected} shows the OM SED data points and their corresponding disc blackbody best-fit models, and also the EPIC-pn two power-law best-fit models of Sect. \ref{xray_mod_sect} for all the 10 \xmm observations. 

So far we have modelled the optical-UV and X-ray data independently of each other; in Sect. \ref{compt_section} we make simultaneous fits to both data sets in order to establish a model that accounts for the broad-band spectrum of Mrk 509 and for its observed variability.

\begin{table}
\caption{Best-fit parameters of the optical-UV continuum, obtained from a disc blackbody model fit to the OM data of each observation as described in Sect. \ref{optical_uv_model_sect}.}
\label{uv_cont_table}
\setlength{\extrarowheight}{3pt}
\begin{minipage}[t]{\hsize}
\renewcommand{\footnoterule}{}
\centering
\begin{tabular}{c | l l l}
\hline \hline
Obs. & $T_{\rm{max}}$ \footnote{Maximum temperature in the disc in eV.} & Norm \footnote{Normalisation area ($A = R_{{\rm{in}}}^2 \cos i$) in $10^{28}\, \mathrm{cm}^{2}$.} & $\chi_\nu ^2$ / d.o.f. \\
\hline

1 & $3.82 \pm 0.06$ & $6.9 \pm 0.6$ & $0.6/11$ \\

2 & $3.89 \pm 0.06$ & $6.7 \pm 0.6$ & $0.6/11$ \\

3 & $3.94 \pm 0.07$ & $6.6 \pm 0.6$ & $0.6/11$ \\

4 & $4.08 \pm 0.08$ & $6.2 \pm 0.6$ & $0.8/11$ \\

5 & $4.15 \pm 0.08$ & $6.0 \pm 0.6$ & $0.8/11$ \\

6 & $4.00 \pm 0.07$ & $6.5 \pm 0.6$ & $0.6/11$ \\

7 & $3.95 \pm 0.07$ & $6.6 \pm 0.6$ & $0.6/11$ \\

8 & $3.86 \pm 0.06$ & $6.9 \pm 0.6$ & $0.6/11$ \\

9 & $3.94 \pm 0.07$ & $6.5 \pm 0.6$ & $0.7/11$ \\

10 & $3.86 \pm 0.06$ & $6.9 \pm 0.6$ & $0.6/11$ \\

\hline
\end{tabular}
\end{minipage}
\end{table}
%

\subsection{Correlation between optical-UV and X-ray count rates}
\label{correlation_sect}
\begin{figure}
\centering
\resizebox{\hsize}{!}{\includegraphics[angle=0]{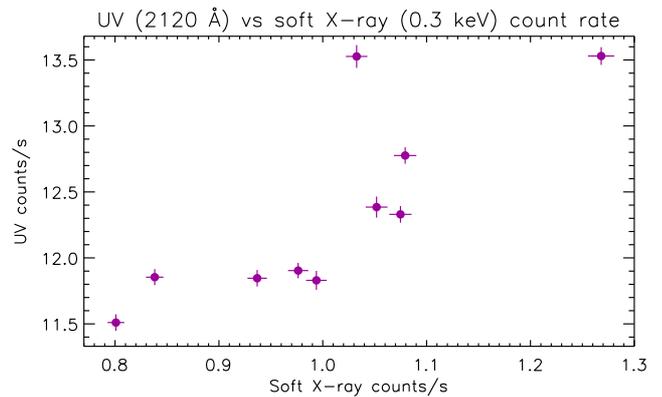}}
\caption{The UV (2120 $\AA$) count rate from the OM UVW2 filter plotted versus the EPIC-pn soft X-ray (0.3 keV) count rate for the 10 \xmm observations. The bandwidth used for the 0.3 keV bin is 0.02 keV.}
\label{uv_soft_xray_fig}
\end{figure}

\begin{figure}
\centering
\resizebox{\hsize}{!}{\includegraphics[angle=0]{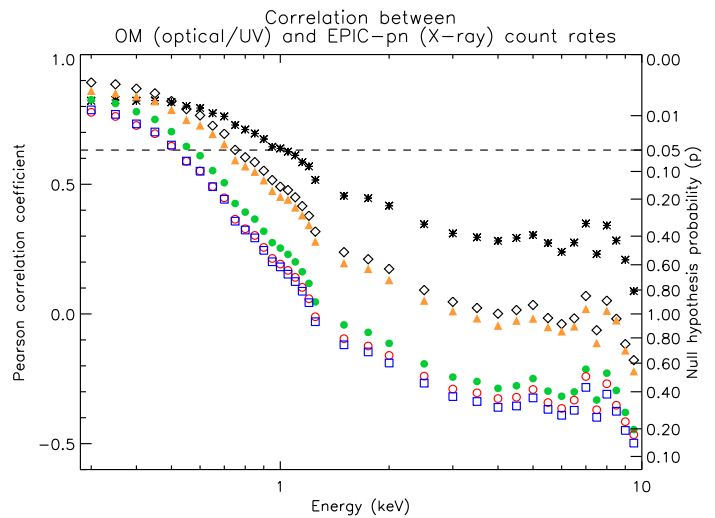}}
\caption{The correlation between the optical/UV count rates from the 6 OM filters and the EPIC-pn X-ray count rate for the 10 \xmm observations at energies from 0.3 keV to 10 keV. Black asterisks: V (5430 $\AA$) filter, black diamonds: B (4500 $\AA$), orange triangles: U (3440 $\AA$), filled green circles: UVW1 (2910 $\AA$), unfilled red circles: UVM2 (2310 $\AA$), blue squares: UVW2 (2120$\AA$). The Pearson correlation coefficients are shown on the left y-axis and the associated null hypothesis probability $p$ (i.e. probability of no correlation) from Student's t-test on the right y-axis. The dashed black line indicates $p$ at 5\%, which shows that below $\sim$ 1 keV the correlation is statistically significant at 95\% confidence.}
\label{uv_xray_correlation}
\end{figure}

Figure \ref{pn_xrt_lightcurve} (blue square data points in the top panel) indicates that the soft X-ray continuum flux of Mrk 509 varied over the 10 \xmm observations in a very similar fashion to the optical-UV continuum (blue square data points in Fig. \ref{uvot_om_lightcurves}). In order to explore the link between the optical-UV and X-ray emission in a model-independent way, we looked at the correlation between their count rates. Fig. \ref{uv_soft_xray_fig} shows the UV (2120 $\AA$) count rate plotted against the soft X-ray (0.3 keV) count rate for the 10 \xmm observations. In fact this correlation is as strong as the one between the UV and optical count rates (the Pearson correlation coefficient between the count rates in the UV and optical filters is about 0.85, which is statistically significant with probability of no correlation, from the Student's t-test, given by $p = 0.002$). Furthermore, the correlation between the optical-UV and X-ray count rates becomes smaller as the X-ray energy gets larger; this is shown in Fig. \ref{uv_xray_correlation}. So the correlation is indeed strongest at those X-ray energies where the soft excess emission is strongest (below $\sim$ 1 keV), implying a strong link between the optical-UV and the soft X-ray excess emission. We investigate this further in the next section by applying a broad-band model to all the data.

\section{Broad-band modelling of the continuum using Comptonisation}
\label{compt_section}

\begin{table*}[!]
\caption{Best-fit parameters of the broad-band warm Comptonisation modelling described in Sect. \ref{compt_section} for each \xmm observation.}
\label{compt_table}
\setlength{\extrarowheight}{3pt}
\begin{minipage}[t]{\hsize}
\renewcommand{\footnoterule}{}
\centering
\begin{tabular}{c | l l | l l l l | l l | l}
\hline \hline
 & \multicolumn{2}{c|}{Disc blackbody:} & \multicolumn{4}{c|}{Comptonisation:} & \multicolumn{2}{c|}{Power-law:} \\ 
Obs. & $T_{\rm{max}}$ \footnote{Maximum temperature in the disc in eV.} & Norm \footnote{Normalisation area ($A = R_{{\rm{in}}}^2 \cos i$) in $10^{29}\, \mathrm{cm}^{2}$.} & $T_\mathrm{seed}$ \footnote{Temperature of the seed photons in eV (coupled to $T_{\rm{max}}$).} & $T_\mathrm{e}$ \footnote{Plasma temperature in keV.}  & $\tau$ \footnote{Optical depth.} & Norm \footnote{$10^{52}$ photons $\mathrm{s}^{-1}$ $\mathrm{keV}^{-1}$ at 1 keV.} & $\Gamma$ \footnote{Photon index.} & Norm $^f$ & $\chi_\nu ^2$ / d.o.f. \\
\hline

1 & $1.96 \pm 0.07$ & $3.9 \pm 0.8$ & 1.96 (c) & $0.21 \pm 0.01$ & $16.5 \pm 0.4$ & $1.1 \pm 0.1$ & $1.89 \pm 0.01$ & $3.7 \pm 0.1$ & $1.24/1796$ \\
2 & $ 2.02 \pm 0.07$ & $3.7 \pm 0.8$ & 2.02 (c) & $0.22 \pm 0.01$ & $16.2 \pm 0.3$ & $1.2 \pm 0.1$ & $1.90 \pm 0.01$ & $3.9 \pm 0.1$ & $1.19/1832$ \\
3 & $2.16 \pm 0.07$ & $3.1 \pm 0.7$ & 2.16 (c) & $0.19 \pm 0.01$ & $17.7 \pm 0.3$ & $1.4 \pm 0.1$ & $1.96 \pm 0.01$ & $4.8 \pm 0.1$ & $1.33/1827$ \\
4 & $2.26 \pm 0.09$ & $2.8 \pm 0.7$ & 2.26 (c) & $0.19 \pm 0.01$ & $17.8 \pm 0.4$ & $1.3 \pm 0.1$ & $1.95 \pm 0.01$ & $3.7 \pm 0.1$ & $1.34/1790$ \\
5 & $2.36 \pm 0.09$ & $2.5 \pm 0.8$ & 2.36 (c) & $0.19 \pm 0.01$ & $18.2 \pm 0.4$ & $1.7 \pm 0.1$ & $1.98 \pm 0.01$ & $4.6 \pm 0.1$ & $1.33/1814$ \\
6 & $2.13 \pm 0.08$ & $3.4 \pm 0.8$ & 2.13 (c) & $0.19 \pm 0.01$ & $17.7 \pm 0.4$ & $1.4 \pm 0.1$ & $1.94 \pm 0.01$ & $4.8 \pm 0.1$ & $1.31/1848$ \\
7 & $2.07 \pm 0.07$ & $3.6 \pm 0.7$ & 2.07 (c) & $0.19 \pm 0.01$ & $17.6 \pm 0.4$ & $1.3 \pm 0.1$ & $1.95 \pm 0.01$ & $4.7 \pm 0.1$ & $1.32/1832$ \\
8 & $2.01 \pm 0.07$ & $3.9 \pm 0.8$ & 2.01 (c) & $0.21 \pm 0.01$ & $16.7 \pm 0.3$ & $1.3 \pm 0.1$ & $1.93 \pm 0.01$ & $4.9 \pm 0.1$ & $1.24/1841$ \\
9 & $2.18 \pm 0.08$ & $2.9 \pm 0.7$ & 2.18 (c) & $0.21 \pm 0.01$ & $16.5 \pm 0.4$ & $1.4 \pm 0.1$ & $1.92 \pm 0.01$ & $4.1 \pm 0.1$ & $1.34/1831$ \\
10 & $1.99 \pm 0.07$ & $3.9 \pm 0.8$ & 1.99 (c) & $0.20 \pm 0.01$ & $17.1 \pm 0.4$ & $1.2 \pm 0.1$ & $1.89 \pm 0.01$ & $4.5 \pm 0.1$ & $1.27/1850$ \\

\hline
\end{tabular}
\end{minipage}
\end{table*}
%

The results of the previous section suggest that the continua in the optical-UV and soft X-ray bands may be physically related, i.e. they may be the results of an underlying process linking the two. Comptonisation up-scattering the optical-UV photons to X-ray energies could be such a process. Thus we applied the {\tt COMT} component in {\sc spex} to fit the optical-UV and X-ray data simultaneously. {\tt COMT} is based on the Comptonisation model of \citet{Tit94}, with improved approximations for the parameter $\beta (\tau)$ which characterises the photon distribution over the number of scatterings which the soft photons undergo before escaping the plasma; for more details see the {\sc spex} manual \citep{SPEX_man}. The seed photons in this component have a Wien law spectrum; we coupled its temperature $T_{\mathrm{seed}}$ to the disc temperature $T_{\mathrm{max}}$ which was left as a free parameter in the broad-band fitting. The up-scattering Comptonising plasma was chosen to have a disc geometry; its parameters are the temperature $T_\mathrm{e}$ and the optical depth $\tau$. The {\tt COMT} component assumes spherical symmetry for the flux calculation. As for our modelling in Sect. \ref{preliminary_sect}, all the absorption and emission corrections discussed in Sect. \ref{corrections_sect}, and the cosmological redshift, were implemented in our modelling here. In addition to the Comptonisation component for modelling the optical-UV continuum and the soft X-ray excess, a power-law component and a Gaussian line were also included to model the hard X-ray continuum and the \FeKa line; the power-law was smoothly broken at low energies before overshooting the optical-UV flux of the disc blackbody and becoming unphysical. In a forthcoming paper on our campaign (Petrucci et al. in prep.), higher energy \integral data will also be included in the broad-band spectral fitting, and the hard X-ray continuum will be modelled by Comptonisation of the disc photons in a hot corona, instead of using a power-law model as done here. In this paper, however, we focus on the direct relation between the disc blackbody emission and the soft X-ray excess and we prefer to use an adequate but simpler representation of the higher energy part of the spectrum. Table \ref{compt_table} shows the best-fit parameters of our broad-band model for each \xmm observation.

\begin{figure}[!]
\centering
\resizebox{\hsize}{!}{\includegraphics[angle=0]{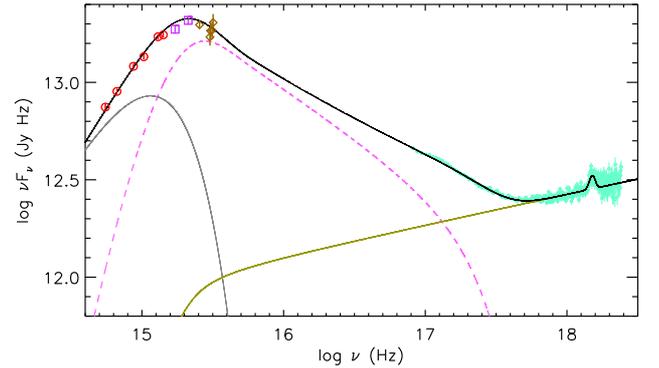}}
\caption{Best-fit broad-band model for \xmm Obs. 2 as described in Sect. \ref{compt_section}. The red circles represent the OM data, the purple squares the \hst/COS data, the brown diamonds are the scaled \fuse data and in green are the absorption-corrected EPIC-pn data. The grey line is the disc blackbody component, the dashed magenta is the warm Comptonisation component and the olive colour line represents the broken power-law and includes a Gaussian for modelling the \FeKa line. The black line is the total model spectrum.}
\label{compt_figure}
\end{figure}

\begin{figure}[!]
\centering
\resizebox{\hsize}{!}{\includegraphics{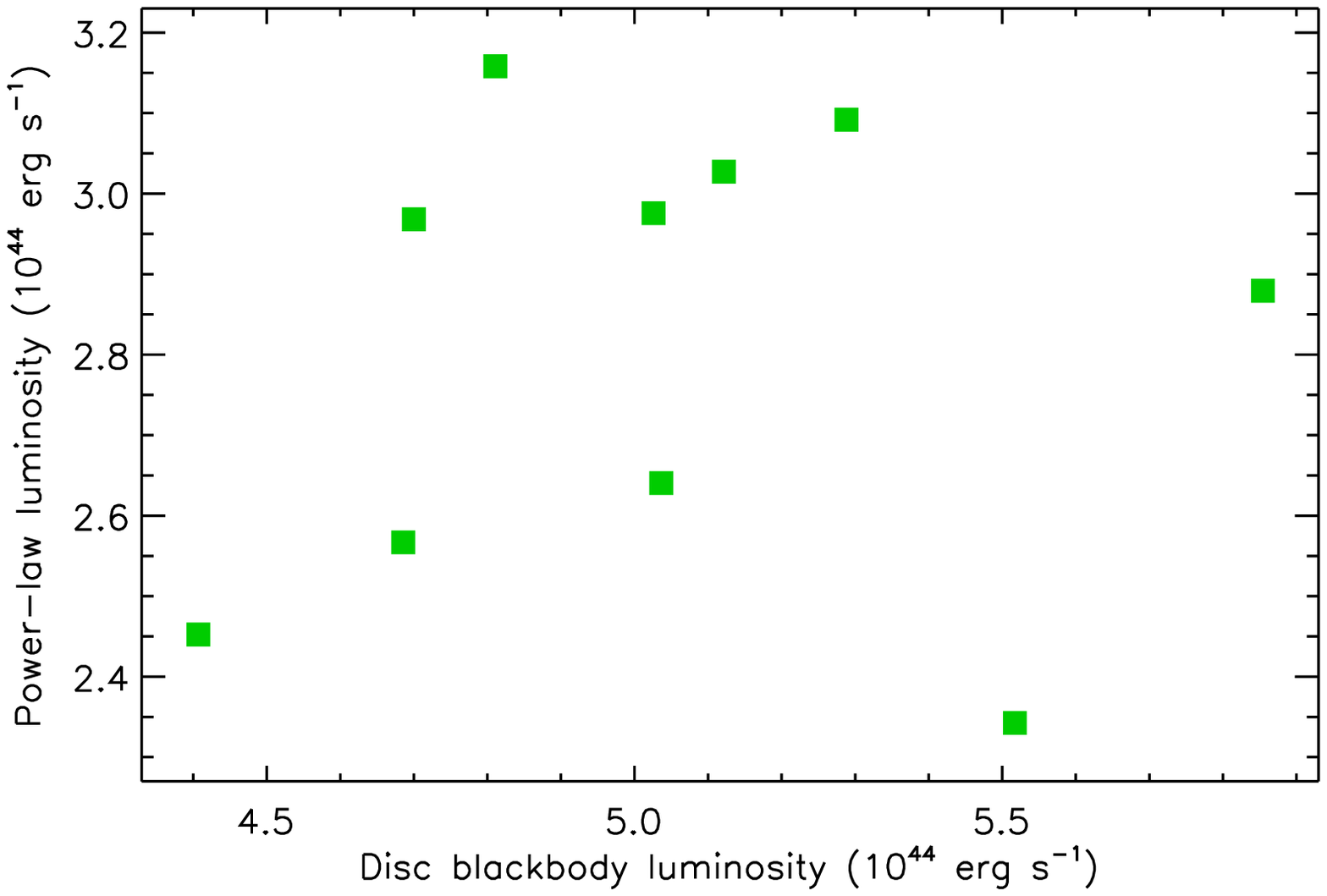}}
\resizebox{\hsize}{!}{\includegraphics{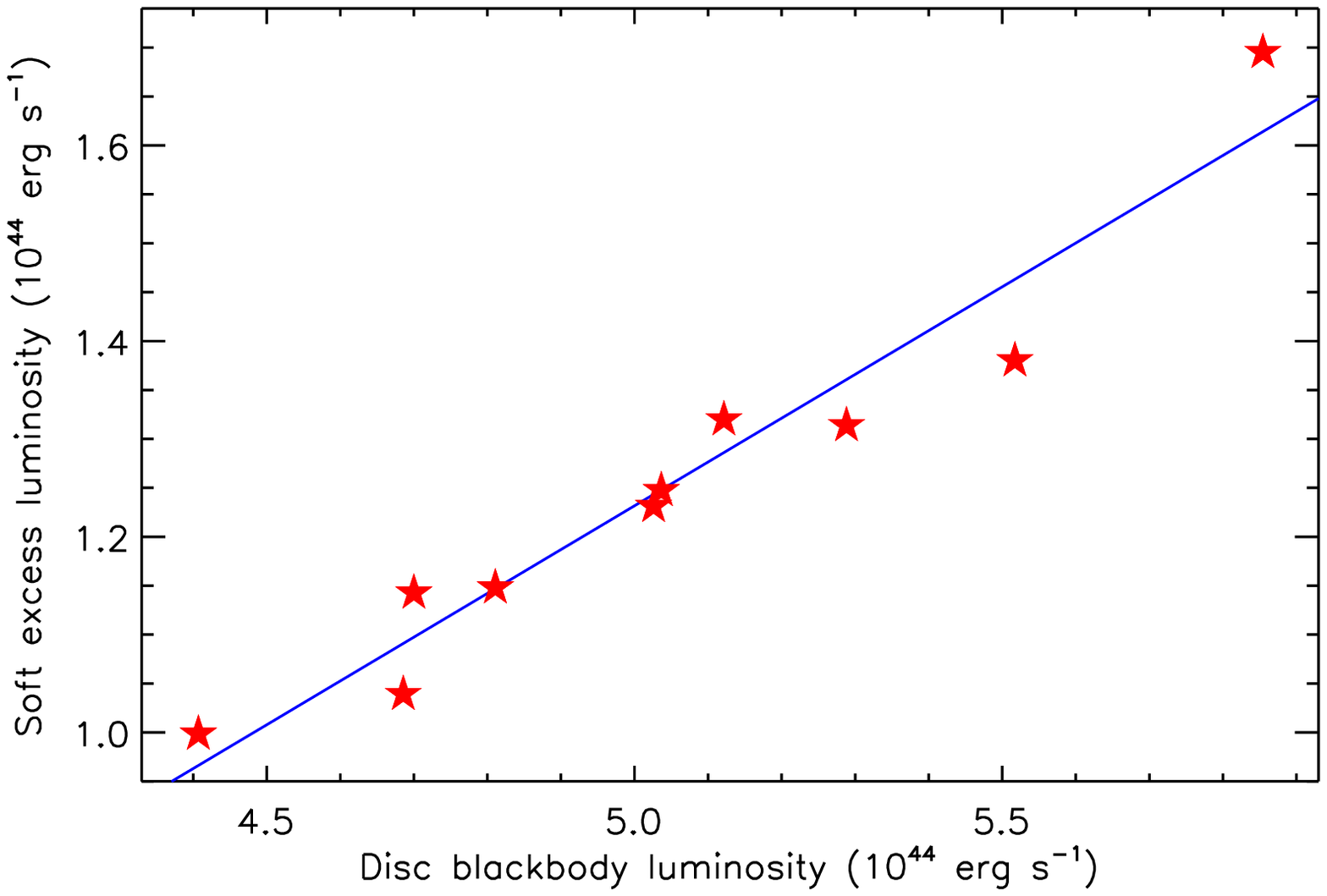}}
\caption{Top panel: the best-fit X-ray power-law luminosity calculated between 0.2--10 keV, plotted versus the disc blackbody bolometric luminosity for the 10 \xmm observations. Bottom panel: 0.2--10 keV luminosity of the soft excess modelled by warm Comptonisation, plotted versus the disc blackbody bolometric luminosity for the 10 \xmm observations. The luminosities are obtained from the broad-band modelling described in Sect. \ref{compt_section}. The blue line in the bottom panel is a fit to the luminosities with linear model: $L_{\rm{soft\ excess}} = 0.448\ L_{\rm disc} - 1.007$, where the luminosities are in units of $10^{44}\ {\rm erg}\ {\rm s}^{-1}$.}
\label{lum_figure}
\end{figure}

Figure \ref{compt_figure} shows the broad-band Comptonisation model for \xmm Obs. 2, displaying individual components contributing to the total model spectrum. Figure \ref{lum_figure} shows luminosity of the X-ray power-law (top panel) and luminosity of the Comptonisation component (bottom panel) plotted versus the disc blackbody bolometric luminosity for the 10 \xmm observations. Interestingly, there is a strong linear correlation between the soft excess modelled by warm Comptonisation and the disc emission, whereas there is no such obvious correlation between the power-law and the disc emission.

%

%
%

\section{Discussion}
\label{discussion}

\subsection{Broad-band continuum variability of Mrk 509}
Optical-UV continuum lightcurves of Mrk 509 (Fig. \ref{uvot_om_lightcurves}) display very similar variability in the 6 filters during the 100 days total duration of the \swift and \xmm campaigns. Specifically, during the 10 \xmm observations, the optical-UV continuum smoothly increased from its minimum in Obs. 1 to its maximum in Obs. 5 (18 days after Obs. 1) and then gradually decreased returning to almost the level of Obs. 1 towards the end of the \xmm campaign. In Sect. \ref{correlation_sect} we showed that there is a strong model-independent correlation between the optical-UV and soft X-ray flux, whereas there is a lack of correlation between the optical-UV and hard X-ray data (see Figs. \ref{uv_soft_xray_fig} and \ref{uv_xray_correlation}). Similarly, using our broad-band model, Fig. \ref{lum_figure} indicates a strong linear correlation between the luminosity of the disc blackbody and that of the Comptonised soft X-ray excess: $L_{\rm{soft\ excess}} = 0.448\ L_{\rm disc} - 1.007$, where the luminosities are in units of $10^{44}\ {\rm erg}\ {\rm s}^{-1}$. However, there is no obvious correlation between the luminosity of the disc and that of the X-ray power-law component. The computed Pearson correlation coefficient between the luminosity of the disc and the Comptonised soft excess component is $+0.97$ (which is statistically significant with {$p=10^{-5}$}), and between the disc and the X-ray power-law is $+0.09$ (which is not significant with {$p= 0.81$}). Also the correlation coefficient between the X-ray luminosity of the Comptonised soft excess and the power-law is $+0.17$ ({$p=0.63$}), which suggests that the fluxes from the two components are not linked over the probed timescales. But there is a significant correlation with coefficient of $+0.88$ ({$p=10^{-3}$}) between the power-law photon index and the disc blackbody luminosity. This behaviour is similar to NGC 7469 \citep{Nan00} in which the UV and hard X-ray fluxes were uncorrelated (like in Mrk 509), but the UV flux and the hard X-ray photon index were correlated. Furthermore, on much larger timescales (few years), \citet{Mar08} have found a correlation between optical and RXTE hard X-ray data. In general, correlations between different energy bands depend on the timescales and resolutions of the monitoring observations.

\subsection{The origin of the soft X-ray excess in Mrk 509}
Here we discuss different interpretations for the origin of the soft X-ray excess, and discuss why Comptonisation is the most likely explanation for its presence in Mrk 509. 

One of the early explanations for the soft X-ray excess emission in AGN was its identification with the high energy tail of thermal emission from the accretion disc (e.g. \citealt{Arn85}, \citealt{Pou86}). As shown in Fig \ref{pn_om_SEDs_corrected}, it is evident that in the case of Mrk 509 the disc blackbody emission from a geometrically thin, optically thick Shakura-Sunyaev disc, cannot extend to the soft X-ray band. Often it is not possible to measure the peak of the disc emission for AGN as it falls in the EUV gap. The temperature of the disc scales with the mass of the black hole as $M^{-1/4}$, and so more massive black holes tend to have lower disc temperatures (see e.g. \citealt{Pet97}). The disc blackbody emission for Mrk 509, which has a relatively high black hole mass of 1--3 $\times 10^8 \ \mathrm{M}_{\odot}$ (see Sect. \ref{accretion_rate_sect}) has a low disc temperature, which causes the peak of the big-blue-bump to be in the detection range of the \hst/COS.

Another interpretation of the soft X-ray excess in AGN is given by \citet{Gie04}, who suggest it could be an artefact of absorption by a relativistically smeared, partially ionised disc wind. They analysed publicly available \xmm EPIC spectra of 26 radio-quiet Palomar-Green (PG) quasars and initially modelled the soft X-ray excess with a Comptonisation component. They found a similar temperature (0.1 to 0.2 keV) for the whole sample, which they claim to be puzzling since the objects in their sample have a large range in mass and luminosity, and hence disc temperature. For this reason they suggest the soft X-ray excess may not be real emission, and propose an alternative solution based on absorption. The relativistically smeared ionised absorption was modelled with {\sc xstar} \citep{Bau01}, convolved with a Gaussian velocity dispersion of $\sim 0.2\, c$. \citet{Gie04} suggest that the smearing is such that it eliminates the possibility of detecting any absorption lines in the RGS spectrum. The absorption creates a wide dip in the EPIC spectrum around 1 keV which results in an apparent soft excess at lower energies. However, we note that by fitting the soft X-ray excess using Comptonisation, there is inevitably large uncertainty in constraining the parameters of the plasma when using only EPIC data as there is degeneracy in the fitting process. In fact, excluding the optical-UV data we can get a good fit to the EPIC data of Mrk 509 with a range of values for the plasma parameters from $T_{\rm{e}} = 0.2\, \rm{keV}$ and $\tau = 18$ to $T_{\rm{e}} = 1.0\, \rm{keV}$ and $\tau = 6$. Therefore, broad-band modelling of simultaneous optical to X-ray data including the peak of the big-blue-bump emission (like performed in this work) is essential to constrain the parameters of the Comptonisation. In most AGN, however, the peak falls in the EUV band where there are no data available, which causes more uncertainty when modelling. The small scatter in the parameters of the plasma found by \citet{Gie04} is likely to be due to the small energy window used for modelling: only using broad-band observations can show the real scatter. Moreover, \citet{Gie04} have strongly constrained their model for all the AGN in their sample by fixing the disc blackbody temperature at 10 eV, the temperature of the hot component at 100 keV and the photon index of the component used to fit the soft X-ray band at 2. These constraints have strong impact on the best-fit results. Furthermore, variability of the soft X-ray excess is crucial in understanding its origin; from our broad-band variability study of Mrk 509, using both EPIC and OM data (together with the \hst/COS and \fuse UV measurements), we have found the soft X-ray to be `real' emission, which varies in a similar fashion to the disc emission.

The soft X-ray excess has also been associated with the relativistically blurred photoionised disc reflection model of \citet{Ros05}. According to this model, the X-ray power-law illuminates a relativistic accretion disc and produces a reflection spectrum which includes a Compton `hump' at about 30 keV, a strong fluorescent \FeKa line at 6.4 keV, and a soft X-ray excess, composed of many emission lines that are blurred by relativistic motion in the accretion disc. \citet{Cru06} have reported that the reflection model makes a better fit to the EPIC-pn spectra of 22 Type-1 PG quasars and 12 Seyfert-1 galaxies than the conventional model of power-law and blackbody. However, no optical-UV data are fitted in \citet{Cru06} and so their reflection modelling is not tested at a broad-band level. Furthermore, fitting the soft X-ray excess using this reflection model requires fine-tuning of the ionisation parameter and inclination angle of the disc, and almost maximally rotating black holes for all the objects. \citet{Don07a} have explored further the disc reflection model to check whether it can account for the soft X-ray excess in AGN, using the additional constraint of hydrostatic balance on the structure of the illuminated disc atmosphere. They conclude that reflection from a hydrostatic disc cannot produce the soft X-ray excess, and constant density disc models require fine-tuning and suppression of the intrinsic continuum to produce the largest observed soft excesses. 

For the case of Mrk 509, we find that the significantly variable part of the X-ray continuum is the soft excess component, which has strong correlation with the variability of the optical-UV disc emission. The hard X-ray power-law displays smaller variability than the soft excess and seems unconnected to the other components over the probed timescales. The reflection spectrum is produced as a result of the illumination of the disc by the power-law and the blurred reflection component is generally expected to be less variable than the power-law. Thus it is unlikely that in Mrk 509 the observed soft X-ray excess, which varies in a very similar fashion to the disc intrinsic emission, is caused by reflection over the probed timescales. In the forthcoming paper by Petrucci et al. (in prep.), where the analysis of our Mrk 509 campaign will be extended to higher energy by the use of \integral data, blurred reflection will be tested in broad-band modelling of the combined \xmm and \integral data sets.

The more likely possibility for the origin of the soft X-ray excess in Mrk 509 is up-scattering of the disc photons in a warm Comptonising corona with lower temperature and higher optical depth than the one responsible for the X-ray power-law (i.e. the hot corona). From broad-band spectral analysis of the Seyfert-1 {NGC 5548}, \citet{Mag98} have found that the soft excess requires a separate continuum component which can be fitted by Comptonisation of thermal photons from a cold disc ($T_{\rm max} \sim 3.2$ eV), in a warm ($\sim 0.3$ keV), optically thick ($\tau \sim 30$) plasma. The authors also find the optical-UV and soft X-ray fluxes in {NGC 5548} to be closely correlated. Their results are similar to those found for Mrk 509 in this work. \citet{Mid09} and \citet{Jin09} also report that the soft X-ray excess in the NLS1s RE J1034+396 and RX J0136.9--3510 can be explained as the result of warm Comptonisation of the disc emission. Furthermore, \citet{Don06} have shown that an optically-thick Comptonising corona over the inner regions of the disc in some Black Hole Binaries (BHBs, e.g. XTE J1550--564) can distort a standard disc spectrum to produce a strong, steep tail extending to higher energies (shown in Figs. 4 and 5 of \citealt{Don06}); this is seen in the Very High State of BHBs, when the X-ray spectra show both a strong disc component and a strong high-energy tail (although in this state of BHBs the hard X-ray power-law seems to be usually absent). The high-energy tail in the Very High State of BHBs may be analogous in nature (although the temperature of the Comptonised gas is substantially higher) to the tail of the warm Comptonisation which appears as a soft excess in the X-ray spectra of Seyferts like Mrk 509. Also, \citet{Zdz01} have fitted the spectrum of the BHB {\object {GRS 1915+105}} in the Gamma-Ray State, by warm Comptonisation of the disc blackbody in a plasma with a temperature of 3.6 keV and an optical depth of 4.4 (see their Fig. 3b). 

Our broad-band variability study of Mrk 509 has shown the existence of a strong link between the disc emission and the soft X-ray excess, suggesting that the disc photons are up-scattered in a warm Comptonising corona close to the inner disc, with a temperature of about 0.2 keV and an optical depth of about 17 (see Table \ref{compt_table}). The scattered fraction of the original disc photons by the corona is $C_{f}[1-\exp(-\tau)]$, where $C_{f}$ is the covering fraction of the corona over the disc and $\tau$ is the optical depth of the corona; this estimation neglects any angle dependencies of the seed and scattered photons. For Mrk 509, $C_{f}$ of the warm corona is estimated to be about 0.25. Furthermore, we calculate that the fraction of accretion power dissipated into the warm corona is also about 0.25.

\subsection{Black hole mass and accretion rate in Mrk 509}
\label{accretion_rate_sect}
From the disc blackbody fits we can estimate the accretion rate $\dot M$ for Mrk 509 using the normalisation ${A = R_{{\rm{in}}}^2 \cos i}$ and the temperature ${T_{*}  = [{3GM\dot M/(8\pi R_{{\rm{in}}}^3 \sigma)}]^{1/4}}$. To do this the mass of the black hole is required. \citet{Pet04} have calculated the black hole mass $M$ of 35 AGN from previously published broad line reverberation-mapping data: ${M = f c \tau_{\rm{cent}} \sigma_{\rm{line}}^{2}/G}$, where ${f = 5.5}$ is the scaling factor, ${\tau_{\rm{cent}}}$ is the emission line time lag relative to the continuum, characterised by the centroid of the cross correlation function, ${\sigma_{\rm{line}}}$ is the width of the emission line, $c$ the speed of light and $G$ the gravitational constant. For Mrk 509 the black hole mass is reported to be $1.43 \times 10^8 \ \mathrm{M}_{\odot}$. However, here we re-calculate the black hole mass using ${\tau_{\rm{cent}}}$ as measured by \citet{Car96}: this is a paper entirely dedicated to the determination of $\tau_{\rm{cent}}$ for the case of Mrk 509 where ${\tau_{\rm{cent}}}$ is found to be 80.2 days for the H$\beta$ line. We also use the H$\beta$ line width measured in this work: ${\sigma_{\rm{line}} = \rm{FWHM}/\sqrt{\ln 256}}$ = 1869 \kms. Adopting these values we calculate a black hole mass of $3.0 \times 10^8 \ \mathrm{M}_{\odot}$ for Mrk 509, which is twice as large as the value reported in \citet{Pet04}. So for a Schwarzschild geometry black hole, from the normalisation $A$ and temperature $T_{*}$, we find $\dot M$ has a range of $0.24$ to $0.34\ \mathrm{M}_{\odot}\ \mathrm{y}^{-1}$ during our \xmm campaign. Assuming a reasonable inclination angle $i$ of $30^{\circ}$ to $45^{\circ}$, we obtain a range of 6.0 to 7.0 $R_{\mathrm{g}}$ for the inner-disc radius $R_{{\rm{in}}}$. 

\section{Conclusions}
\label{conclusions}

In order to study the intrinsic optical-UV and X-ray continuum of the Seyfert-1/QSO hybrid Mrk 509 (using \xmm OM and EPIC-pn, \swift UVOT and XRT, \hst/COS and \fuse), we have taken into account various processes occurring in our line of sight towards the central source, such as Galactic reddening and absorption, host galaxy emission, BLR and NLR emission and the AGN multi-phase warm absorption. We conclude that:

\begin{enumerate}

\item The optical-UV continuum of Mrk 509 is consistent with thermal emission from a geometrically thin, optically thick cold accretion disc with a maximum disc temperature of about 2 eV. The X-ray continuum can be represented by two distinct components: a power-law with $\Gamma \sim 1.9$ and a soft X-ray excess below 2 keV.  

\item We have investigated the variability of the optical-UV and X-ray continuum over the 100 days duration of our monitoring campaign with a resolution of a few days. We have found that the variability of the soft X-ray excess is very similar in profile to that of the optical-UV emission from the disc. The flux variability of the higher energy X-ray power-law component (albeit smaller), on the other hand, is uncorrelated with the variability of the soft X-ray excess and that of the disc emission over the probed timescales. There is however a correlation between the power-law photon index and the disc blackbody flux.

\item We can account for the origin of the soft X-ray excess and its variability by Comptonisation of the disc photons in a warm (about 0.2 keV), optically thick ($\tau \sim 17$) corona surrounding the inner disc. The X-ray power-law is likely to originate from Comptonisation in a separate optically-thin hot corona, possibly located further out, in the form of an atmosphere surrounding the disc.

\item Explaining the soft X-ray excess using warm Comptonisation has previously been suggested for the Seyfert-1 {NGC 5548} and the NLS1s: {RE J1034+396} and {RX J0136.9-3510}. This interpretation is analogous to the inner-disc warm-corona model which explains the existence of the high-energy tail observed in some BHBs spectra in the Very High State. Mrk 509 is the highest mass black hole system known to display such variability and origin for the soft excess. Only variability studies based on multi-wavelength monitoring campaigns, such as the one reported in this paper, allow the detailed investigations that can reveal the processes leading to the formation of AGN broad-band spectra.

\end{enumerate}

\begin{acknowledgements}
This work is based on observations obtained with \xmm, an ESA science mission with instruments and contributions directly funded by ESA member states and the USA (NASA). This work made use of data supplied by the UK Swift Science Data Centre at the University of Leicester. SRON is supported financially by NWO, the Netherlands Organization for Scientific Research. MM acknowledges the support of a PhD studentship awarded by the UK Science \& Technology Facilities Council (STFC). POP acknowledges financial support from the French space agency CNES and the French GDR PCHE. GAK gratefully acknowledges support from NASA/XMM-Newton Guest Investigator grant NNX09AR01G. Support for HST Program number 12022 was provided by NASA through a grant from the Space Telescope Science Institute, which is operated by the Association of Universities for Research in Astronomy, Inc., under NASA contract NAS5-26555. GP acknowledges support via an EU Marie Curie Intra-European Fellowship under contract no. FP7-PEOPLE-2009-IEF-254279. AJB acknowledges the support of a STFC Postdoctoral Fellowship. MC acknowledges financial support from contract ASI-INAF n. I/088/06/0. KCS acknowledges the support of Comit\'e Mixto ESO - Gobierno de Chile. MM acknowledges useful discussion with R. Soria, V. Yershov and M. J. Page. Finally, we thank the anonymous referee whose useful suggestions and comments improved the paper.
\end{acknowledgements}

\end{document}